\documentclass[prl,preprintnumbers,tightenlines,twocolumn,superscriptaddress]{revtex4-1}

\usepackage{amsmath}
\usepackage{amssymb}
\usepackage{graphicx}
\usepackage{psfrag}
\usepackage{color}
\usepackage[utf8]{inputenc}
\usepackage{ifthen}
\usepackage{listings}
\usepackage{placeins}

\newcommand{\bra}[1]{\langle #1 |}

\newcommand{\ket}[1]{| #1 \rangle}

\newcommand{\abs}[1]{\left\vert #1 \right\vert}

\newcommand{\E}[1][\empty]{
  \ifthenelse{\equal{#1}{\empty}}
    {\mathbb{E}}
    {\mathbb{E}\left( #1 \right)}
}
\renewcommand{\exp}[1][\empty]{
  \ifthenelse{\equal{#1}{\empty}}
    {\mathrm{exp}}
    {\mathrm{e}^{#1}}
}

\newcommand{\psit}[1][\empty]{%
  \ifthenelse{\equal{#1}{\empty}}
    {\psi_t}
    {\psi_t^{(#1)}}
}

\newcommand{\npsit}[1][\empty]{%
  \ifthenelse{\equal{#1}{\empty}}
    {\tilde\psi_t}
    {\tilde\psi_t^{(#1)}}
}

\newcommand{\SI}{Supplemental Material}
  
\usepackage{ulem}  
\graphicspath{{{./}}}
\begin{document}
\title{Excitonic Wave Function Reconstruction from Near-Field Spectra Using Machine Learning Techniques }

\author{Fulu Zheng}
\affiliation{Max-Planck-Institut f\"ur Physik komplexer Systeme, N\"othnitzer Strasse\ 38, 
D-01187 Dresden, Germany }
	
\author{Xing Gao}
\affiliation{Max-Planck-Institut f\"ur Physik komplexer Systeme, N\"othnitzer Strasse\ 38, 
D-01187 Dresden, Germany }
\affiliation{Department of Chemistry,
University of Michigan, Ann Arbor, 48109-1055, Michigan, USA}
	
\author{Alexander Eisfeld}
\email{eisfeld@pks.mpg.de}
\affiliation{Max-Planck-Institut f\"ur Physik komplexer Systeme, N\"othnitzer Strasse\ 38,
D-01187 Dresden, Germany }

\begin{abstract}
 A general problem in quantum mechanics is the reconstruction of eigenstate wave functions from measured data. In the case of molecular aggregates, information about excitonic eigenstates is vitally important to understand their optical and transport properties. 
 Here we show that from spatially resolved near field spectra it is possible to reconstruct the underlying delocalized aggregate eigenfunctions. Although this high-dimensional nonlinear  problem defies standard numerical or analytical approaches, we have found that it can be solved using a convolutional neural network. For both one-dimensional and two-dimensional aggregates we find that the reconstruction is robust to various types of disorder and noise.
\end{abstract}
\maketitle

Self-assembled molecular aggregates on dielectric surfaces  are not only promising candidates for optoelectronic devices \cite{Goronzy2018, Hoffmann-Vogel2018, Schmaltz2015,Gaberle2017, Casalini2017}, but they are also paradigmatic systems to study collective molecular properties. 
The surface degrees of freedom barely couple to molecular excitations and fluorescence is only weakly quenched \cite{Goronzy2018, Muller2013, Gebauer2004}.
Strong interactions between the transition dipoles of the molecules lead to delocalized excitonic eigenstates where an electronic excitation is {\it coherently} shared by many molecules \cite{,Hoffmann2000,Eisfeld2017, Muller2013_JCP}.
 Knowledge of these excitonic eigenstates is crucial for understanding the optical and transfer properties of the aggregates. 
 However, these states are difficult to probe, in particular since most states are typically inaccessible with far field radiation, due to selection rules.

Ways to circumvent far-field selection rules have been discussed  for various systems \cite{Takase2013, Iida2011, Jain2012}, and it has been  shown theoretically  for the case of molecular aggregates that essentially all eigenstates become optically accessible when exposed to an electromagnetic field that is {\it spatially inhomogeneous} on the  scale of a few ($\sim 10$) nanometers  \cite{Gao2018}.  
There are many possibilities to realize  such near fields experimentally and to detect the optical response \cite{Kim2015, Chen2016, Berweger2012, Neacsu2010, Gramotnev2014, Zhang2010, Becker2016,RaAlZe18_26365_,EsBeWi19_698_};
an overview is given in Refs.~[\onlinecite{HeGo18_365_}] and [\onlinecite{ChHuMe19_1804774_}].

\begin{figure}
\includegraphics[width=8.5cm]{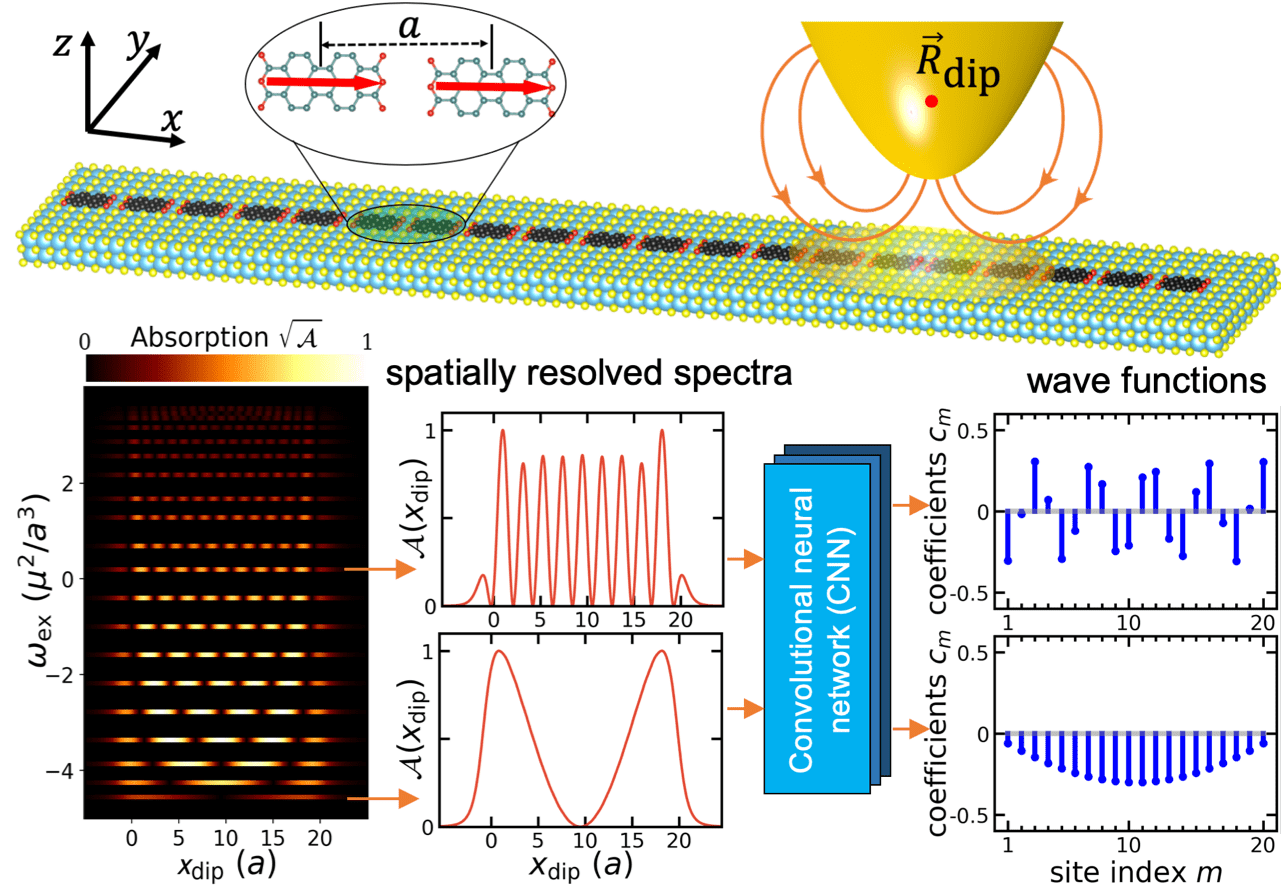}
\caption{Top: Sketch of the setup. The red arrows represent the transition dipole moments $\vec{\mu}$ of the molecules. We use distance  $a$ as our unit of length.
The aggregate is located in the $x$-$y$ plane. 
Bottom: Left: spatially and frequency resolved near-field spectrum stemming from a Hertzian dipole located $2a$ above the aggregate  and with dipole pointing along the $z$ axis. Middle: spatially resolved spectra for fixed frequency are fed into a CNN. Right: extracted wave functions corresponding to the respective spectra.  
}
\label{fig:setup}
\end{figure} 
We consider the setup proposed in Ref.~\cite{Gao2018}, which is sketched  in the upper panel of Fig.~\ref{fig:setup}. 
At the apex of a metallic tip the electromagnetic field for optical excitation is created. 
A beneficial feature of the setup is the possibility to scan the tip across the aggregate, obtaining  an topographical image similar to that of an atomic force microscope, and also recording  a frequency dependent optical spectrum with high resolution at each tip position. 
 By considering slices at the frequencies where one has absorption maxima, one finds the position dependent absorption corresponding to the respective eigenstate. 
 Complicating this process is the nonlinear dependence of the spectra on the eigenstates; see Eq.~(\ref{eq:abs}). 
 A mapping from the spectra to the eigenstates has so far not been found, and hence a fundamental question is this: 
 do these slices contain enough information to reconstruct the corresponding wave function?
  Furthermore, how can the inversion be performed in practice?

Here we  show that both questions can be answered in the affirmative.
To this end we use a numerical approach based on machine learning.
Machine learning techniques have penetrated many areas in material science, chemistry, and physics \cite{MeBuWa19_1_,carleo2019machine}, and have been applied to various problems including the extraction of
parameters from spectroscopic data \cite{Rodriguez2019, Borodinov2019, Giri2019} and the estimation of quantum states \cite{Torlai2018, Carrasquilla2019}.
Many different machine learning methods exist.
In this Letter, we use convolutional neural networks (CNNs), famous for image classification \cite{KrSuHi12_1097_}, to construct the mapping from the spectra to the eigenstates.
Because of the huge amount of possible wave functions we use a regression based approach instead of the more common classification approach. 
 
 Although CNNs are extremely versatile, for each physical problem the appropriate architecture must be selected and a suitable set of training data obtained. This data set must be very large and contain many diverse examples, which we generate using a theoretical model that closely mimics the experimental conditions. 
 Once the training data are collated, we tailor the choice of architecture to it. 


For  modeling of the aggregate we use a widely adopted description\cite{May2011,Gao2018}. 
For each of the $N$ individual molecules comprising the aggregate we take two electronic states into account: the ground state $\ket{g}_n$ and the first excited state $\ket{e}_n$, where the index $n$ labels the monomers which are placed at position $\vec{R}_{n}$. 
The transition dipole between these two states is denoted by $\vec{\mu}_n$. 
Initially the aggregate is in the global ground state $\ket{g_{\rm agg}}=\ket{g}_1\cdots \ket{g}_N$.  
For linear absorption we are interested in states with one excitation. Using basis states $\ket{m}=\ket{e}_m\prod_{n\neq m}^{N}\ket{g}_n$ the excited state Hamiltonian for the system is written as
\begin{align}
\label{eq:Hamiltonian}
 H_{\rm ex}=\sum_{m=1}^N\varepsilon_m\ket{m}\bra{m}+\sum_m\sum_{n\neq m}V_{mn}\ket{m}\bra{n}.
\end{align}
Here $\varepsilon_m$ is the transition energy of molecule $m$ and $V_{mn}$ is the transition dipole-dipole interaction, which we take as  
$
V_{mn}=\big[\vec{\mu}_m\cdot \vec{\mu}_n - 3 (\vec{\mu}_m \cdot \vec{u}_{mn})(\vec{\mu}_n \cdot \vec{u}_{mn})\big] / R_{mn}^3
$
with $\vec{R}_{mn}$ the distance vector from molecule $m$ to $n$, $\vec{u}_{mn} = \vec{R}_{mn} / R_{mn}$ and $R_{mn}=|\vec{R}_{mn}|$. 
From the  Schr\"odinger equation $H_\text{ex}\ket{\phi}=E\ket{\phi}$ one obtains the $N$ eigenenergies $E_\ell$ with corresponding eigenstates 
 \begin{equation}
 \label{eq:Wave-function}
 \ket{\phi^{(\ell)}}=\sum_{m=1}^{N}c_{m}^{(\ell)}\ket{m}. 
 \end{equation}

Tips suitable for the present purpose can at present be fabricated with apex diameters of a few nanometers and can be placed with high accuracy at distances of $\sim\, 1 \, \mathrm{nm}$ above the investigated object \cite{EsBeWi19_698_}.
The excitation field could be created by the radiation of a  laser beam focused at the apex region \cite{ChHuMe19_1804774_} or by a localized surface plasmon polariton \cite{Berweger2012,JiChLi18_881_,EsBeWi19_698_}.
As for the case of far field absorption of molecular aggregates on dielectric surfaces, the emitted intensity (which to a good approximation is proportional to the number of absorbed photons) is detected \cite{MLeSc11_241203_,Eisfeld2017}.
In most cases of interest there is a ``bright'' state  providing the required far-field emission.

We model the electromagnetic field stemming from the tip as a Hertzian dipole located at position $\vec{R}_{\rm dip}$ a few nanometers (tip distance + radius of the tip) above the aggregate.
The resulting field has a strong spatial variation on the scale of a few nanometers in the plane of the aggregate.
With $\vec{E}(\vec{R}_m;\vec{R}_{\rm dip})$ we denote the electric field component at the position of monomer $m$. 
More details can be found in Eq.~(S2) of the {\SI} \cite{SI}.
The absorption strength from the ground state to the eigenstate $\ket{\phi^{(\ell)}}$  can be written as \cite{Gao2018}
	\begin{equation}
	\label{eq:abs}
\mathcal{A}^{(\ell)}(\vec{R}_{\rm dip})= \abs{\sum_{m=1}^{N}c_{m}^{(\ell)}\ \vec{\mu}_{m}\cdot\vec{E}(\vec{R}_m; \vec{R}_{\rm dip})}^2.
	\end{equation}
Equation~(\ref{eq:abs}) is valid for fields that vary strongly over the extent of the aggregate but only very weakly over the extent of a single molecule.
For brevity we often drop the eigenstate-index $\ell$ and write $c_m$ for the wave function coefficients and  $\mathcal{A}(\vec{R}_{\rm dip})$ for the corresponding absorption strength.


To reconstruct the eigenstate wave functions Eq.~(\ref{eq:Wave-function}) from the near-field absorption spectra we use CNNs with several convolutional layers.  
Since we only intend to show the feasibility of the inversion, we do not optimize the architectures.
The inputs are spatially discretized near-field spectra $\mathcal{A} (\vec{R}_{\rm dip})$ evaluated at a finite number $N_{\rm tip}$ of  positions $\vec{R}_{\rm dip}$ at fixed distance from the aggregate.
The input for the CNN is therefore a real-valued array with $N_{\rm tip}$ elements.
 The outputs are the corresponding predicted coefficients $c_{m}^{\rm pre}$  of the wave function, which we have normalized to fulfill $\sum_m |c_{m}^{\rm pre}|^2=1$. 
The dimension of the output is equal to the number of molecules $N$. 
To evaluate the quality of the prediction of the CNN we use
the ``loss function''
$
L=\frac{1}{4}\sum_{m=1}^{N} |c_{m}-{c_{m}^{\rm pre}}|^2 ,
$
 where $c_m-{c_{m}^{\rm pre}}$ is the difference between true and predicted coefficients. 
 This loss function, which takes values between 0 and 0.5, is minimized to optimize the parameters inside the CNN during the training.

Adequate network training requires a sufficient amount of appropriate training data, which in our case is pairs of wave functions and corresponding spatial spectra. 
Na\"{i}vely, one might choose random wave functions and  the corresponding spectra.
However, these random wave functions will typically not be ``close'' to eigenfunctions of the aggregate and consequently the  network trained on this data will not be able to predict the physical eigenstates.
Therefore, we use a physically motivated procedure to construct a viable data set for training.
We calculate the wave functions by diagonalizing the Hamiltonian (\ref{eq:Hamiltonian}) for random values of the parameters $\epsilon_n$ and $V_{nm}$,  physically corresponding to static disorder typically caused by ``defects'' of the surface.

In the following we show results for the case of a one-dimensional (1D) chain of $N=20$ molecules (with an arrangement shown in Fig.~\ref{fig:setup}) and a two-dimensional (2D) arrangement with $N=N_x\times N_y= 10 \times 5$ molecules (the positions and orientations of the molecules are sketched in the bottom panel of Fig.~\ref{fig:examples}). 
These arrangements are motivated by the experimental structures of PTCDA on KCl \cite{Muller2013_JCP,Eisfeld2017}.
In the results shown in this Letter we use a distance $z_{\rm dip}=2\,a$ between sample and dipole, which for the PTCDA case corresponds roughly to $3\,\mathrm{nm}$.
We use CNNs consisting of several convolutional layers, having a  large number (several hundred thousand) of trainable parameters. 
The training data are constructed as follows.
The fluctuations of the site energies are obtained from a normal distribution $\delta \epsilon_n\sim \mathcal{N}(0,\sigma_{\rm d}^2)$ where we choose various values of the variance $\sigma_{\rm d}^2$ to sample a broad range of wave functions. 
For the 2D case  we use $\sigma_{\rm d}=0.02,\ 0.04,\ 0.06,\ 0.08$, and we generate $2000$ realizations of each disorder strength. 
All disorder values are in units of the maximal dipole-dipole interaction  in the aggregate and are consistent with the estimates from the experimental data of Refs.~\cite{Muller2013_JCP,Eisfeld2017}.
For each realization we have $N=50$ eigenstates resulting in a training set of size 400,000.  
For the 1D case we include even larger disorder (up to $\sigma_{\rm d}=5$) as well as disorder in the dipole-dipole couplings to cover an even larger wave function space containing strongly localized wave functions, giving altogether $\sim 4.6$ million samples.
Examples of wave functions and corresponding spectra are presented in Fig.~\ref{fig:examples}. 
To emphasize the variety of wave functions that we cover, in the \SI{} \cite{SI} we have included  additional examples and also more details about the used $\sigma_{\rm d}$.
There we also provide details of the used network architectures and the training process.

\begin{figure}[pt]
\hspace{-0.2cm}
\includegraphics[width=8.cm]{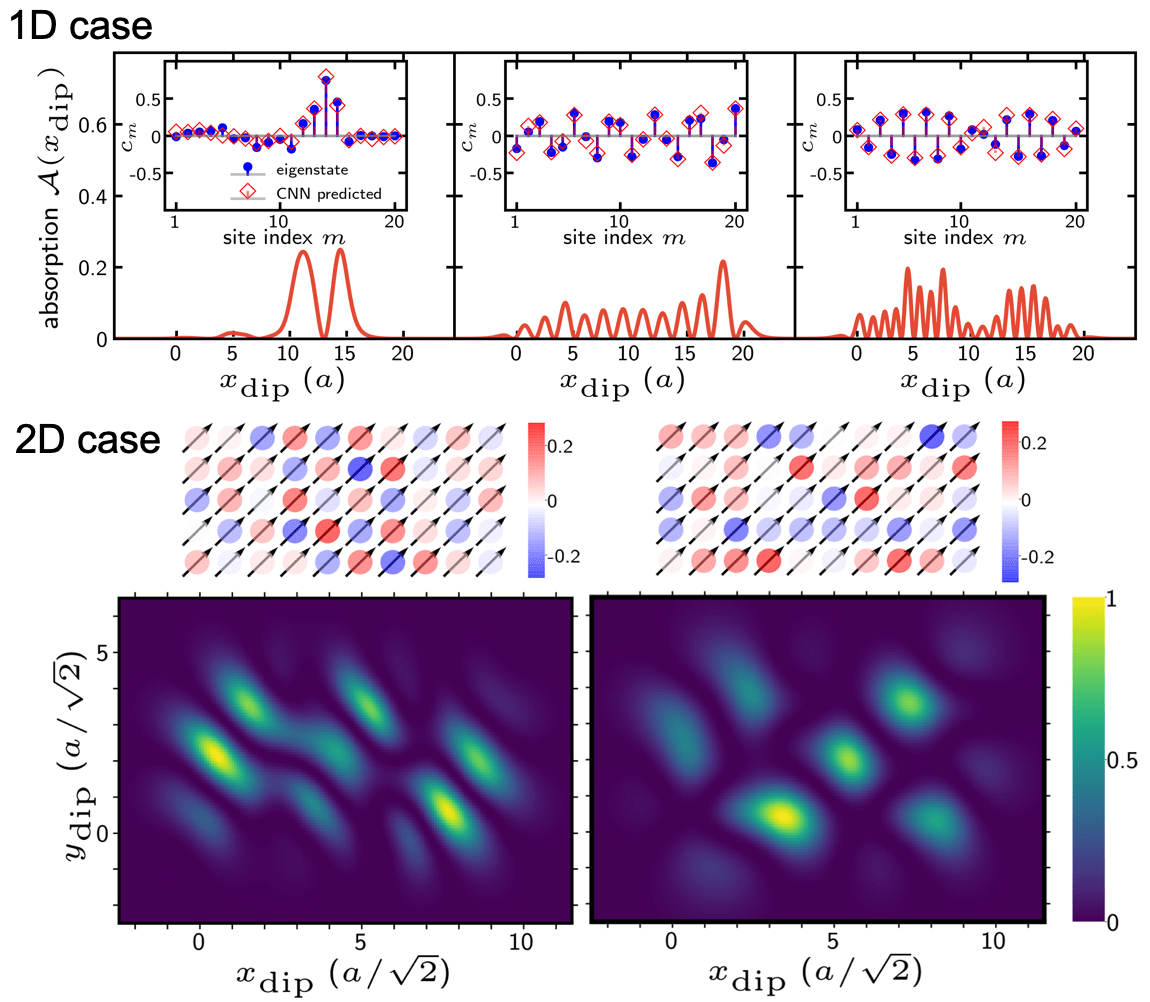}
\caption{Examples of eigenstate wave functions and corresponding spectra $\mathcal{A}(\vec{R}_{\rm dip})$ used to train the CNN for a 1D linear chain with $N=20$ molecules (top) and a 2D array with $N=10\times 5$ molecules (bottom). 
The excitation-dipole is located at a distance $z_{\rm dip}=2a$ above the aggregate (the distance $a$ is defined in Fig.~\ref{fig:setup}).   
The wave functions corresponding to the spectra are shown in the insets in the 1D case (together with the predictions of the CNN) and in the upper panel in the 2D case (the arrows indicate the orientation of the molecular transition dipoles). 
In all cases shown here the loss is $10^{-2}$. 
Further examples, including the disorder-free case, can be found in the \SI{} \cite{SI}.}  
\label{fig:examples}
\end{figure}

Training leads for both cases quickly to low values of the loss around $10^{-2}$ (see Figs.~S1b and S3b of the \SI \cite{SI}).
To help provide some intuition for this loss value, we chose the examples presented in  Fig.~\ref{fig:examples}  such that their loss value is exactly $10^{-2}$. 
 One sees that for values on the order of $10^{-2}$ the agreement between predicted and true wave functions is very good.
 More examples can be found in the \SI \cite{SI}, where we present in particular all reconstructed wave functions for the disorder-free case.
 For the disorder free case we have, even for the high lying states, with rapidly varying wave function phase, a loss-value smaller than $10^{-3}$, i.e.,\ perfect agreement.

The CNNs not only give correct predictions for the training (and validation) data, but also correctly predict ``unseen'' test data.
We have generated many additional spectra diagonalizing Hamiltonian (\ref{eq:Hamiltonian}) for disorder strengths $\sigma_{\rm d}$ not used during training. 
The performance of our CNNs on these test-data sets is summarized in Fig.~\ref{fig:LossAnalyzation}.
For the most relevant case of small diagonal disorder ($\sigma_{\rm d}\lesssim 0.05$) the average loss-value is well below $10^{-2}$, indicating very good reconstruction of the coefficients. 
Even for larger values of the disorder the CNNs perform quite well (note that in the 2D case  the value $\sigma_{\rm d}\lesssim 0.1$ is even larger than the largest disorder in the training data set). 


%
\begin{figure}[tp]
\includegraphics[width=8.5cm]{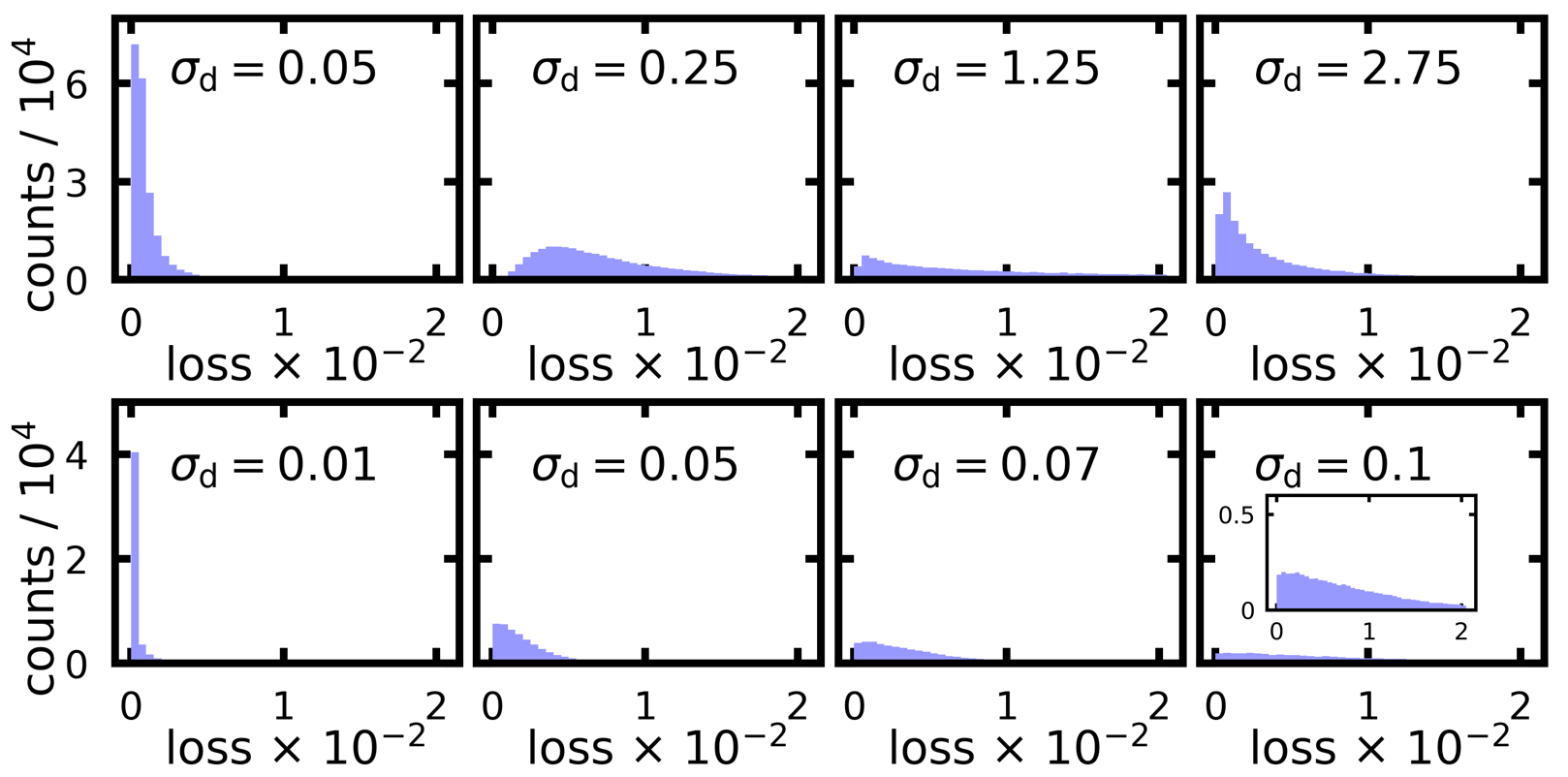}
\caption{\label{fig:LossAnalyzation} Distribution of loss-values for different values of $\sigma_{\rm d}$ as indicated in each panel for the 1D (top) and the 2D (bottom) cases. For each $\sigma_{\rm d}$ the number of realizations is $10\,000$ (1D case) and $1000$ (2D case). We take all eigenstates into account. 
}
\end{figure}
%

So far, we have discussed an ideal situation:  the aggregate is  described by the simple Hamiltonian (\ref{eq:Hamiltonian}) and we assume a perfect measurement of the absorption strength with high spatial resolution.
In the following we show that even in more general cases one can use CNNs to obtain information about the underlying wave functions from the spatial spectra. We focus in the following on the 1D case because the smaller size of the problem allows us to obtain better statistics.

We first consider the case of lower spatial resolution. 
In the examples shown above the spatial resolution was around $0.08\,a$, which corresponds roughly to 1\,\AA\, for a molecular distance $a\approx 1.25\,\mathrm{nm}$.  
We have found that even for a resolution that is four times lower we can perfectly reconstruct the wave functions (see Figs.~S10 and S11 of the {\SI} \cite{SI}).

We now turn to the situation when there is noise on the measured spectra. 
Such noise could stem from detector noise or imperfect tip positioning.
We mimic such noise by simply adding independent random numbers to the absorption values for each tip position.
The strength of the noise $\sigma_\mathrm{n}$ is for each spectrum taken relative to the maximal absorption  $\mathrm{max}[\mathcal{A}(\vec{R}_\mathrm{dip})]$; thus the random numbers are drawn from a Gaussian distribution where the standard deviation is  $\sigma_\mathrm{n} \mathrm{max}[\mathcal{A}(\vec{R}_\mathrm{dip})]$.
Examples of such noisy spectra are shown in the bottom row of Fig.~\ref{fig:LossAnalyzationNoise} for $\sigma_{\rm n}=0.03$, $0.1$, and $0.2$; further examples can be found in Section III\,C of the \SI{} \cite{SI}.
The CNN trained without noise was unable to predict the wave functions in this case, and so we retrained the CNN using the same training and validation data as before, but with added noise with $\sigma_{\rm n}=0.1$.
Remarkably, then the network is not only able to correctly predict the wave functions for the same noise strength, but it also works for the noise-free case  and even noisier cases, as demonstrated in Fig.~\ref{fig:LossAnalyzationNoise}.
One sees in the upper row of that figure  that even for $\sigma_\mathrm{n} = 0.2$ the loss values are well below $10^{-2}$.
The spectra, in the bottom row  are examples of a spatially resolved spectrum for a noise realization with the $\sigma_\mathrm{n}$ used in the upper row.

\begin{figure}[tp]
\includegraphics[width=8.0cm]{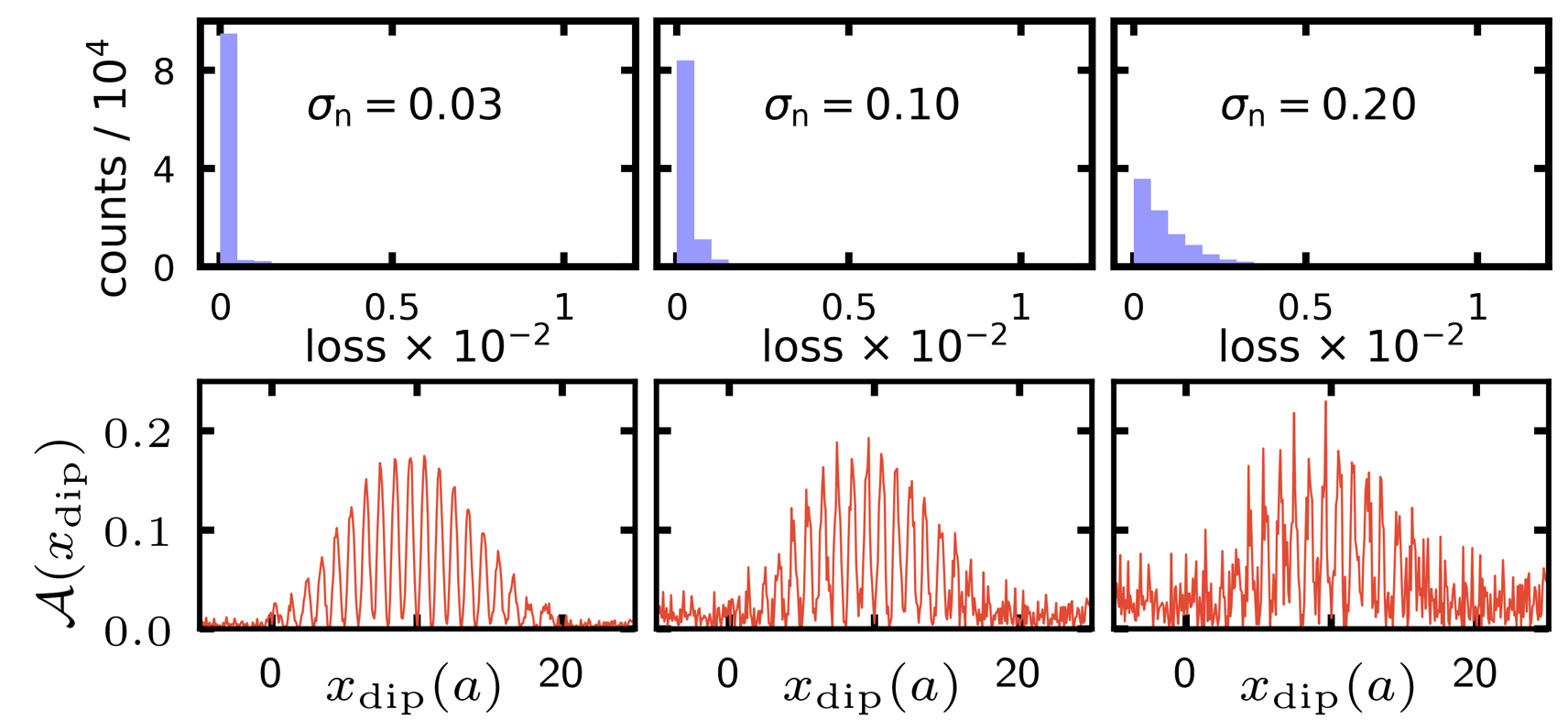}
\caption{\label{fig:LossAnalyzationNoise}
Top: Distribution of the loss-values when noise is added to the spectra of the 1D disorder-free system. 
In each panel a specific value $\sigma_\mathrm{n}$ for the relative strength of the noise has been used, which is provided in the panel. 
For all cases the CNN has been trained with $\sigma_{\rm n}=0.1$.  
Bottom: For the $\sigma_{\rm n}$ values of the top row we show one example of how the noise alters the spectrum. In all these cases the same noise free spectrum is used.
}
\end{figure}

In modeling the aggregate we have made the approximation that Hamiltonian (\ref{eq:Hamiltonian}) does  not contain coupling to additional dynamical degrees of freedom which cause broadening of the absorption lines.
Such broadening can stem for example from internal molecular vibrations or from vibrations of the surface.
In the far-field experiments of PTCDA on KCl this broadening has been estimated to be  in the order of a few wave numbers (at temperatures around $10\,$K) while the typical energy scale of the transition dipole-dipole interaction is at least an order of magnitude larger \cite{Muller2013_JCP,Eisfeld2017}.
An additional broadening mechanism is induced by the tip itself, which due to its polarizability can interact with the molecular transition dipoles in a similar way as the molecules interact with each other \cite{Gao2018}.
For tip diameters around $5\,\mathrm{nm}$ this interaction is quite small, but increases quickly with increasing diameter.
As long as the environmental influence is small, a wave function based description is still useful. 
To assess the influence of the above mentioned mechanisms on the predicted coefficients, we have calculated near-field spectra taking molecular line broadening and the interaction with the tip into account using the local-field theory approach of Ref.~\cite{Gao2018}.
We find that even for line broadening slightly smaller than the energy separation between eigenstates the reconstructed coefficients are  nearly the same as in the situation without line broadening.
For realistic material parameters of the tip and a tip diameters $<10\,\mathrm{nm}$ the coupling of the molecular transitions to the tip plasmon is weak and does not strongly influence the reconstructed coefficients. 
Note that, with increasing tip diameter, also the distance of the excitation dipole to the sample increases, which implies a loss of resolution and a decrease of the signal intensity in particular for fast oscillating wave functions. 
Nevertheless, in our examples even for diameters around $10\,\mathrm{nm}$ the reconstruction of the wave function works still well.
Details of our findings are presented in Section III\,D of the \SI{} \cite{SI}. 
The above considerations show that a reasonable reconstruction of the eigenfunctions is feasible not only in an ideal situation, but also under realistic experimental conditions.
For a specific experimental implementation one would calculate the training data using parameters as close as possible to the respective experimental situation.

 For the aggregates that we have considered we could easily diagonalize millions of Hamiltonians. 
 For larger two-dimensional aggregates (number of molecules $N>100$) one would want to diagonalize  fewer Hamiltonians. 
 One way of generating enough training data is to perform only a few diagonalizations of the disordered Hamiltonian without noise, and then simply add noise to the so obtained wave-function coefficients.
  Since for large aggregates the input and output of the CNN have larger sizes, one probably also has to use a slightly more optimized architecture than the ones we have used in the present Letter.

To conclude, we have shown that properly constructed CNNs can obtain wave functions from their spectra to a very high accuracy even in the presence of significant experimental noise and imperfect conditions, thus confirming that the inversion of Eq.~(\ref{eq:abs}) is possible and providing a method for doing so. 
We have tried to obtain the wave-function coefficients by using various optimizers, including Gaussian process regression (for details see Section IV of the {\SI} \cite{SI}). 
We only obtained reasonable results for quite small aggregates ($N\lesssim 10$) and wave functions with a small number of nodes, cementing the need for sophisticated machine learning techniques.

In addition to the application to near-field spectra of molecular aggregates, the methodology adopted in the present Letter can be applied to spectra of other systems, such as those of electron energy-loss spectroscopy of silver nanowires \cite{Rothe2019} where a similar inversion as Eq.~(\ref{eq:abs}) is required. 
Hence the implementation of CNNs, as shown here, opens new possibilities to extract key insight from experimental spectra.

\begin{acknowledgments}
A. E. acknowledges support from the DFG via a Heisenberg fellowship (Grant No. EI 872/5-1). We thank M.T. Eiles for many helpful comments.
\end{acknowledgments}


 %

\widetext
\clearpage

\begin{center}
\textbf{\large -- Supplemental Material -- \\Excitonic Wave Function Reconstruction from Near-Field Spectra Using Machine Learning Techniques }
\end{center}
\setcounter{equation}{0}
\setcounter{figure}{0}
\setcounter{table}{0}
\setcounter{page}{1}
\makeatletter
\renewcommand{\theequation}{S\arabic{equation}}
\renewcommand{\thefigure}{S\arabic{figure}}
\renewcommand{\bibnumfmt}[1]{[S#1]}
\renewcommand{\citenumfont}[1]{S#1}

\renewcommand\thepage{S\arabic{page}}
\newcommand{\RNum}[1]{\uppercase\expandafter{\romannumeral #1\relax}}

\graphicspath{{{./}}}
\title{-- Supplemental Material -- \\Excitonic Wave Function Reconstruction from Near-Field Spectra Using Machine Learning Techniques }

\author{Fulu Zheng}
\affiliation{Max-Planck-Institut f\"ur Physik komplexer Systeme, N\"othnitzer Strasse\ 38, 
D-01187 Dresden, Germany }
	
\author{Xing Gao}
\affiliation{Max-Planck-Institut f\"ur Physik komplexer Systeme, N\"othnitzer Strasse\ 38, 
D-01187 Dresden, Germany }
\affiliation{Department of Chemistry,
University of Michigan, Ann Arbor, 48109-1055, Michigan, USA}
	
\author{Alexander Eisfeld}
\email{eisfeld@pks.mpg.de}
\affiliation{Max-Planck-Institut f\"ur Physik komplexer Systeme, N\"othnitzer Strasse\ 38,
D-01187 Dresden, Germany }
\maketitle

\vspace{1cm}
In this Supplemental Material we provide additional details on the following items:
\begin{itemize}
\item[\ref{Hertz}.] The spatially varying light field of a Hertz dipole.
\item[\ref{sec:2Dcase}.] The 2D aggregate. Architecture and training procedures of the convolutional neural network (CNN). Examples of spectra and predicted wave functions.
\item[\ref{sec:1Dcase}.] Information regarding the 1D case
\begin{itemize}
\item[\ref{sec:Prediction_examples}.] Wave function prediction for the disorder free case and for various disorder realizations.
\item[\ref{sec:low_res}.]  Wave function reconstruction using the CNN trained using spectra with lower spatial resolution.
\item[\ref{sec:noise}.] Training and wave function prediction for the case with noise added to the spectra.
\item[\ref{sec:polarizability}.] Homogeneous broadening and finite tip size.
\end{itemize}
\item[\ref{sec:GaussianProcess}.]  Eigenstate wave function reconstruction from the near-field spectra with Gaussian Process Regression.
\end{itemize}

\clearpage
\newpage
\section{\label{Hertz}The spatially varying light field of a Hertz dipole}

We consider electromagnetic radiation of frequency $\omega$ with an electric field component that at time $t$ and position $\vec{r}$ given by
\begin{align}
  \label{eq:elecric_field}
    \vec{E}(\vec{r},t)=\text{Re}\Big\{\vec{E}(\vec{r})e^{i\omega t}\Big\}.
\end{align}
Note the explicit dependence of the electric field on the position $\vec{r}$.

In the Letter a Hertzian dipole with dipole moment $\vec{d}$, located at $\vec{R}_{\rm dip}$ creates the electromagnetic field. In the near field zone this field can be written as 
\begin{align}
  \label{eq:hertz_near}
  \vec{E}(\vec{R}_m;\vec{R}_{\rm dip})
  = \frac{1}{4\pi\epsilon_{0}} \Bigg\{ \dfrac{3(\vec{R}_m-\vec{R}_{\rm dip})\Big[(\vec{R}_m-\vec{R}_{\rm dip})\cdot\vec{d}\Big]}{|\vec{R}_m-\vec{R}_{\rm dip}|^5} -\dfrac{\vec{d}}{|\vec{R}_m-\vec{R}_{\rm dip}|^3} \Bigg\}.
\end{align}
Here $\vec{R}_m$ denotes the position of molecule $m$. We have here explicitly included the factor $1/4 \pi \epsilon_0$ which contains the dielectric constant $\epsilon_{0}$. 
\clearpage
\newpage

\section{\label{sec:2Dcase}The two-dimensional (2D) case}

\noindent
The CNNs are constructed and trained using Keras \cite{chollet2015keras_SI} (version 2.2.4) with the optimizer {\em Adam} \cite{Kingma2015_SI}.

\begin{figure}[hpb]
 \begin{minipage}[t]{8cm}
    \includegraphics[width=8cm]{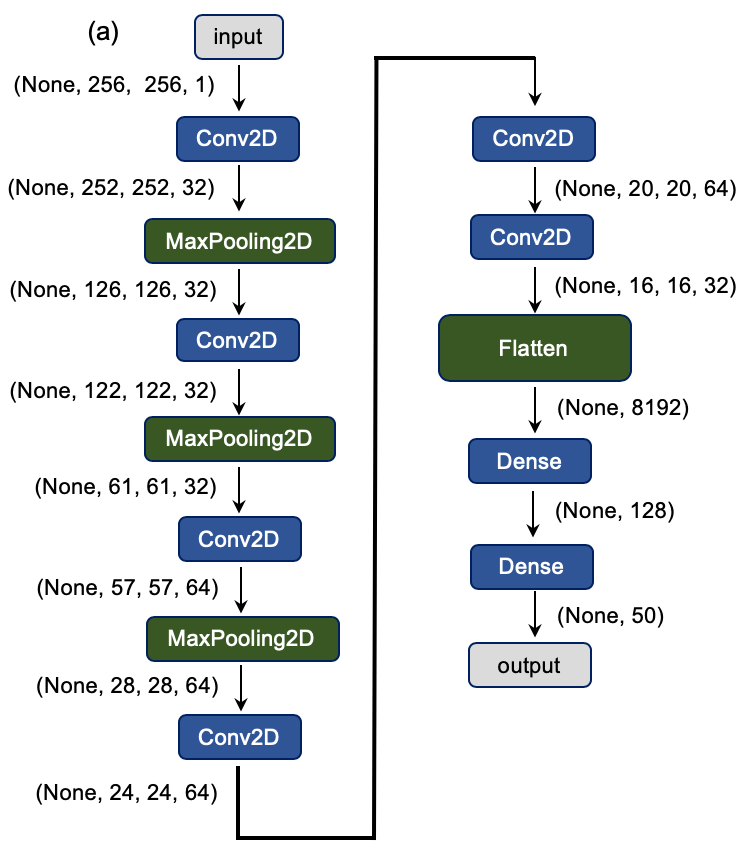}
 \end{minipage}
 \begin{minipage}[b]{8cm}
    \includegraphics[width=7.4cm]{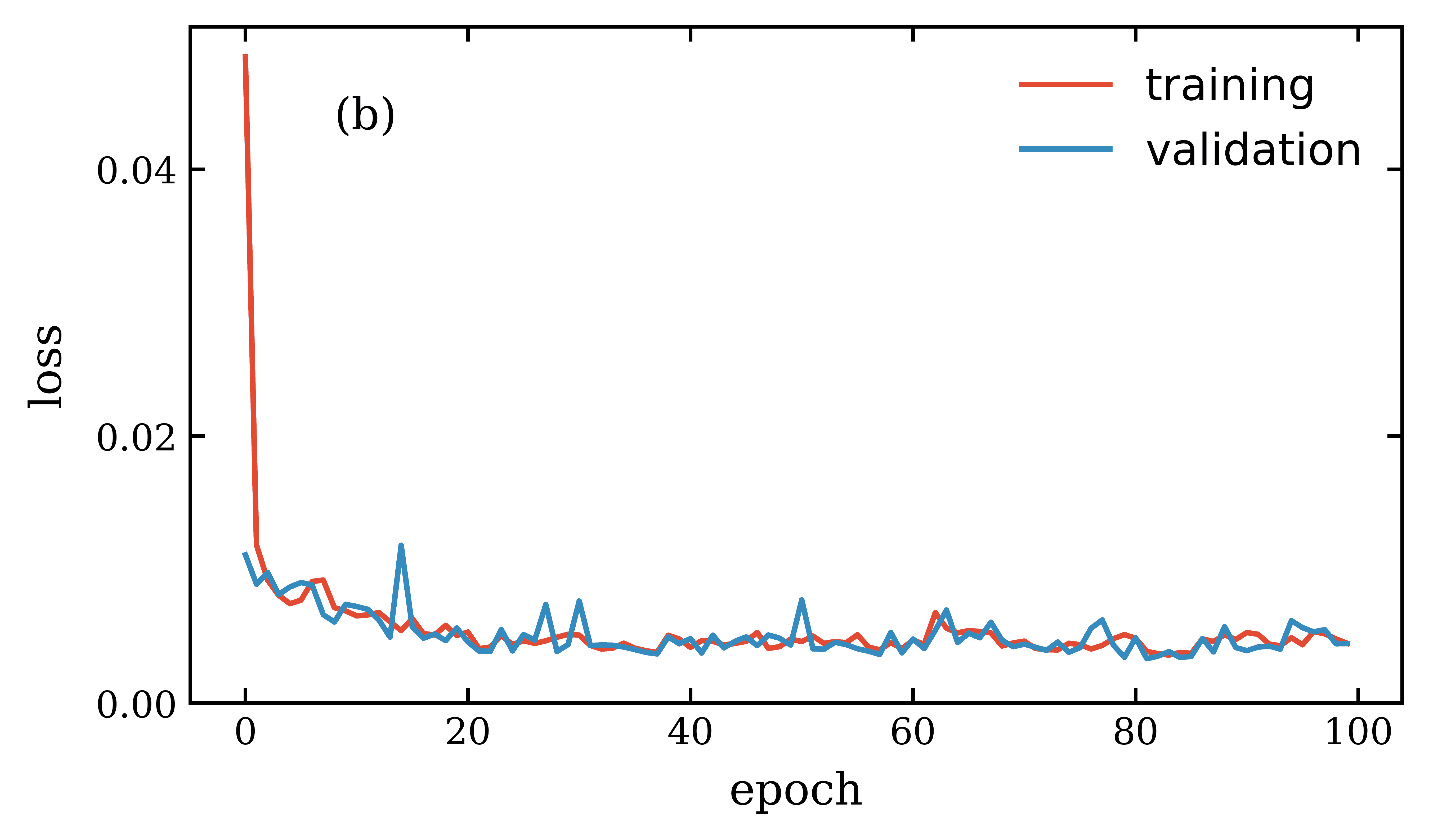}
    \includegraphics[width=7cm]{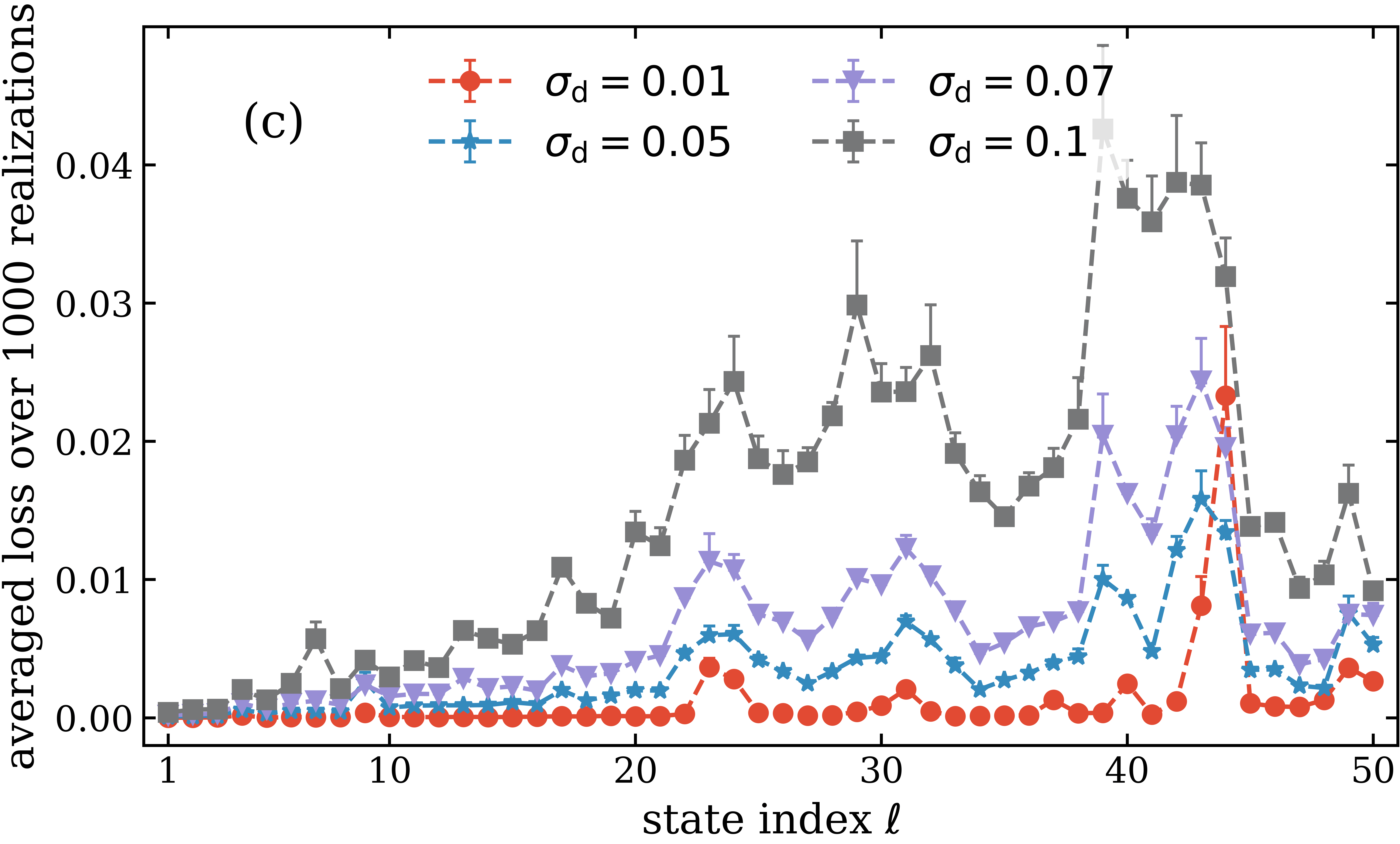}
 \end{minipage}
\caption{\label{fig:Architecture2D}(A) Architecture of the CNN used for the 2D aggregates described in the Letter. The output shape of each layer is shown in brackets. (b) Training and validation loss of the CNN for 2D array trained for 100 epochs. (c) Quality of the prediction for ``unseen'' test data. For each eigenstate index $\ell$ the mean loss (averaged over 1000 realization with disorder strength $\sigma_\mathrm{d}$) is plotted. The eigenstates are labeled according to increasing energy.}
\end{figure}

 \vspace{0.3cm}
 \noindent
{\it Architecture:} 
The architecture of the CNN is illustrated in Fig.~\ref{fig:Architecture2D} (a). In total the network has $\sim 1\,300\,000$ trainable parameters.  

\begin{figure}
  \begin{minipage}{8.5cm}
    no disorder  ($\sigma_{\rm d}=0$)\\
    \includegraphics[width=8cm]{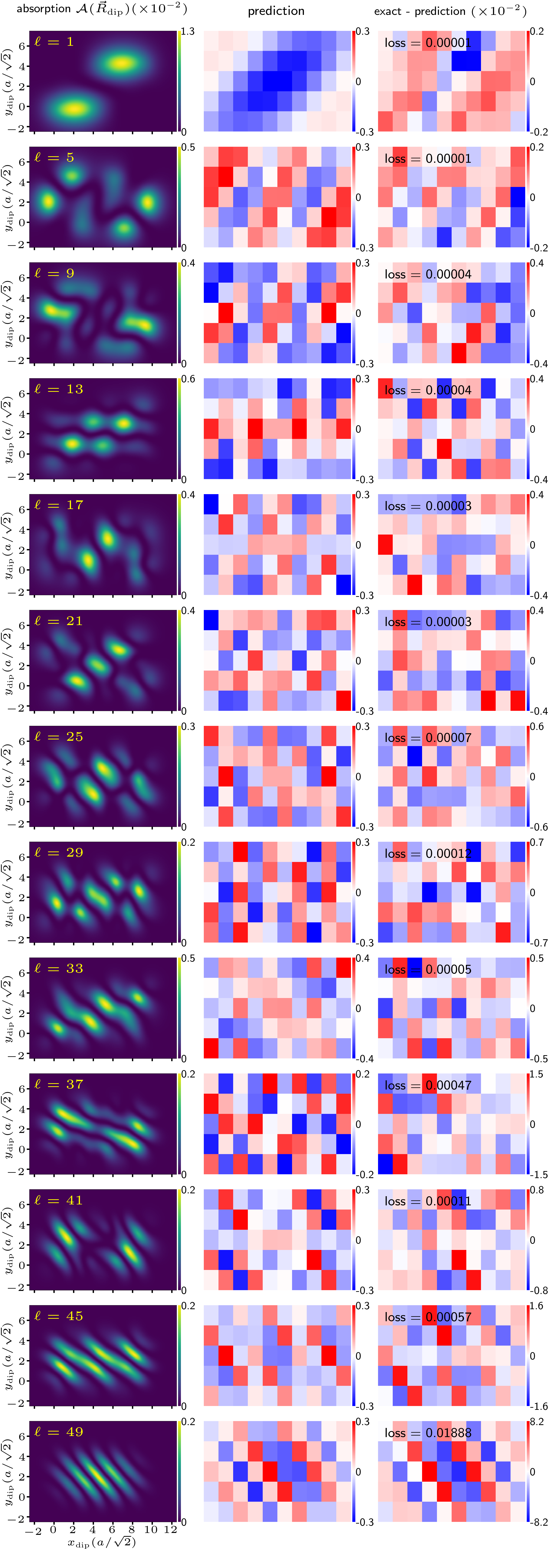}
  \end{minipage}
  \begin{minipage}{8.5cm}
    with disorder $\sigma_{\rm d}=0.1$
    \includegraphics[width=8cm]{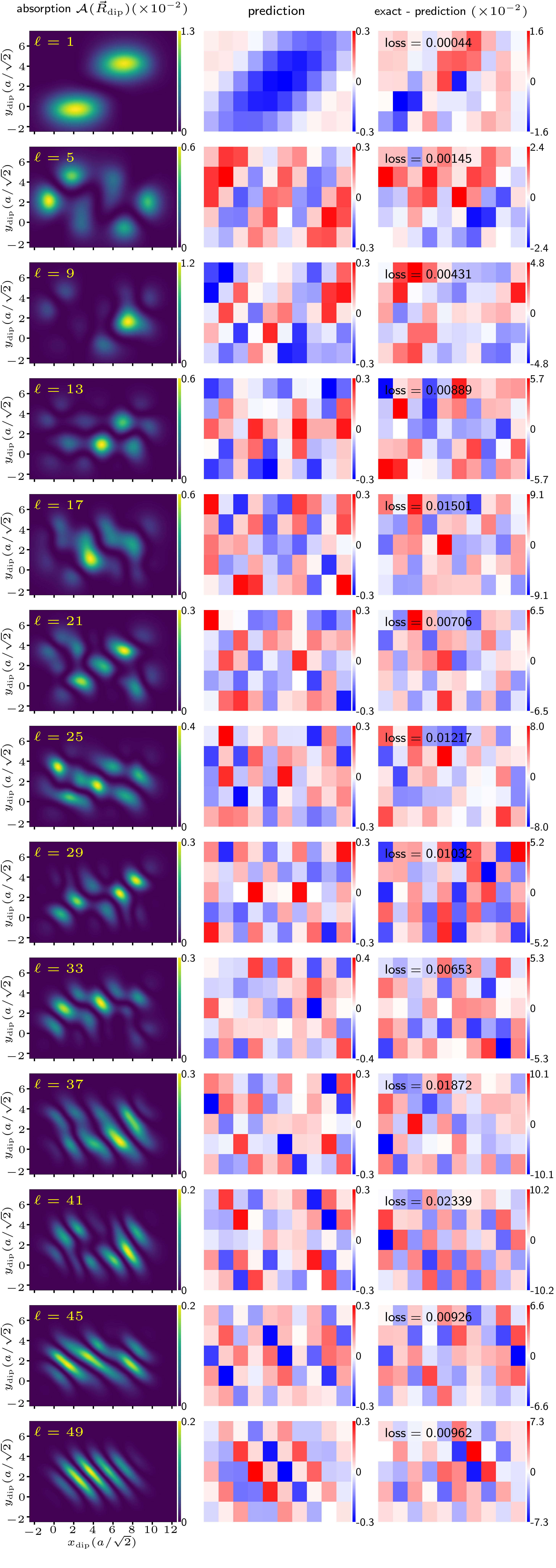}
    \end{minipage}
\caption{\label{fig:Examples_2D}Examples of spatially resolved absorption spectra and corresponding predicted wave functions in the 2D-aggregate case. Different panels show typical examples for different disorder values $\sigma_{\rm d}$ (provided in each panel). The left panel is the disorder free case. In each panel, the left row shows spectra for selected eigenstates. The eigenstate label $\ell$ is given in each plot (we label states according to increasing energy). In the middle column the corresponding wave functions predicted by the CNN are shown. In the right column the difference between predicted wave function and exact wave function is given. In each plot also the corresponding loss value is provided.}
\end{figure}

 \vspace{0.3cm}
 \noindent
 {\it Training data:} As described in the Letter, we add disorder $\delta \varepsilon_m \sim \mathcal N (0, \sigma_{\rm d}^2)$ to the molecular site energy $\varepsilon_m$ to generate the training data. Here  $\mathcal N (0, \sigma_{\rm d}^2)$ denotes a Gaussian distribution with zero mean and standard deviation $\sigma_{\rm d}$. For the results of the Letter we used  $\sigma_{\rm d} \in \{0.02,\ 0.04,\ 0.06,\ 0.08\}$. 
 For each of these disorder strengths, we considered 2000 realizations. This leads to $4\times 2000 =8000$ different Hamiltonians that we diagonalize. Since the considered 2D-aggregate has $10\times 5 =50$ molecules from each Hamiltonian we obtain $50$ wave functions. That means we have  $8000 \times 50 = 400\,000$ training data samples in total. Examples of the wave functions and the respective spatial near field spectra are shown in Fig.~2 of the Letter. In Fig.~\ref{fig:Examples_2D} we show further examples, that illustrate the dependence on the eigenstate label and on the disorder. We label the eigenstates according to increasing energy and the spatially resolved spectra are evaluated at $N_{\rm dip} = 256 \times 256 = 65\,536$ tip positions.

\vspace{0.3cm}
\noindent
{\it Validation data:} The validation data, that is used during training, is produced similar to the training data, but now taking $500$ realizations for each disorder strength.

\vspace{0.3cm}
\noindent
\textit{Training:} We trained the CNN for 100 epochs with a batch size of 64. The training and validation loss during training is shown in Fig.~\ref{fig:Architecture2D} (b).

\vspace{0.3cm}
\noindent
{\it Test:} As discussed in the Letter, testing was performed on ``unseen'' samples generated from disorder with $\sigma_\mathrm{d}$ values ($\sigma_\mathrm{d} = 0.01, 0.05, 0.07$, and $0.1$) not used during training. In Fig.~3 of the Letter the distribution of loss is shown when considering 1000 realizations for each $\sigma_\mathrm{d}$ value. In Fig.~\ref{fig:Architecture2D} we present this date in a slightly different way. First we label for each realization the eigenenergies according to increasing energy. Then for each eigenstate label $\ell$ we average the loss values belonging to this label. These values are plotted in Fig.~\ref{fig:Architecture2D}(c).

\vspace{1cm}

\clearpage
\newpage

\section{\label{sec:1Dcase} The one-dimensional (1D) case:}
\begin{figure*}[phb]
\centering
  \begin{minipage}[t]{8.cm}
    \includegraphics[width=7.cm]{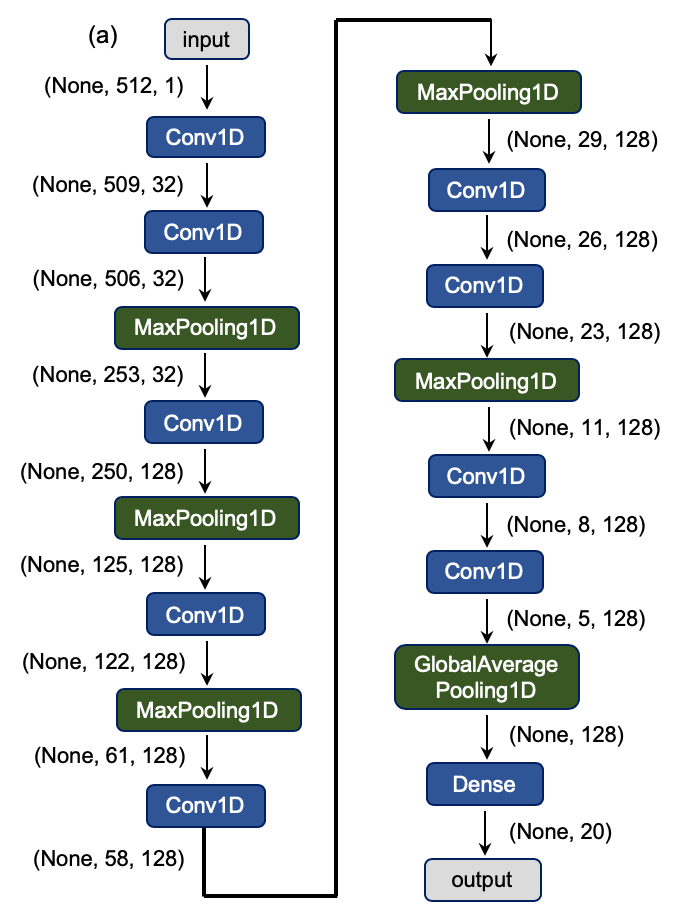}
  \end{minipage}
  \begin{minipage}[b]{8cm}
    \includegraphics[width=7.7cm]{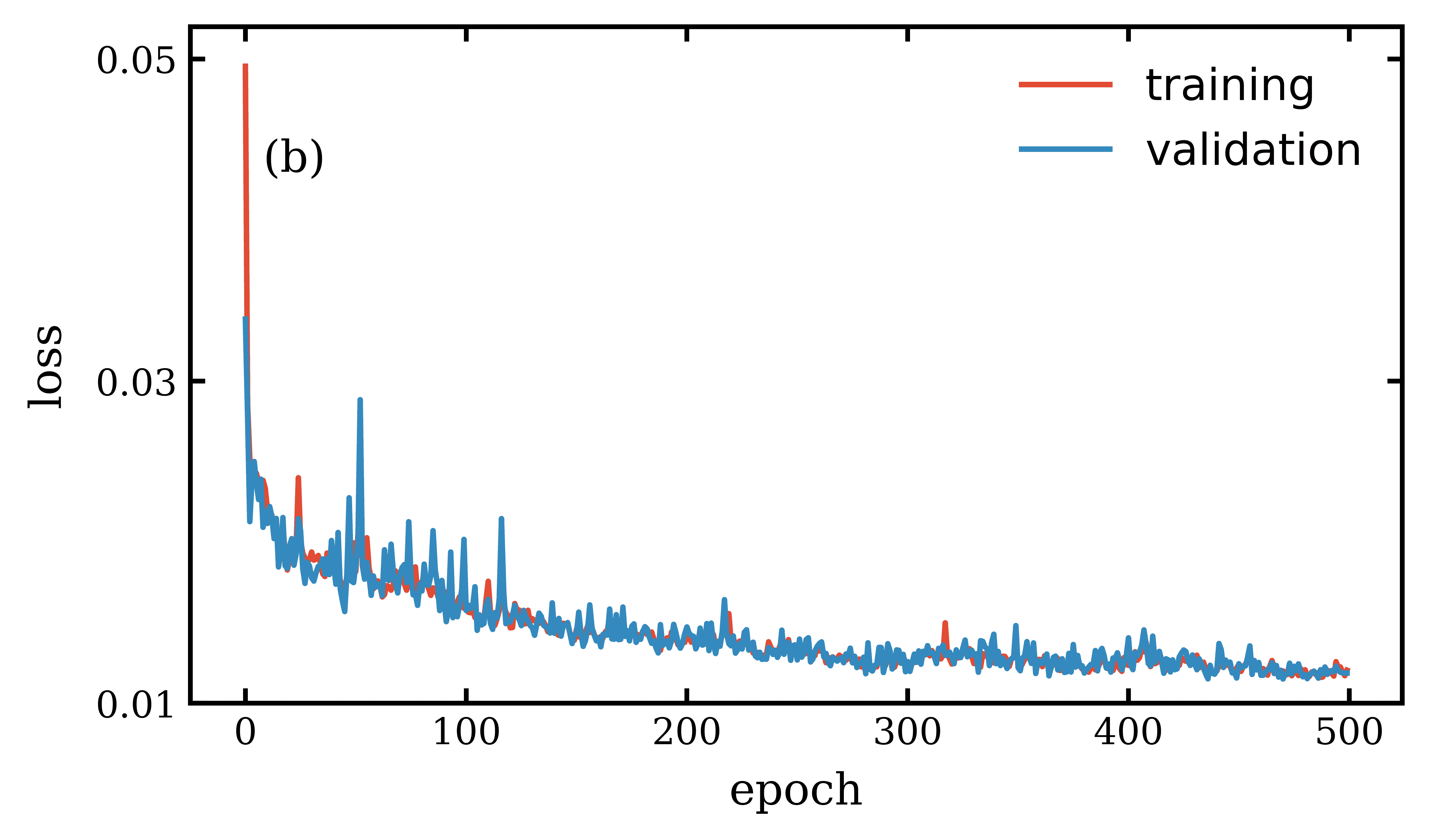}\\
    \includegraphics[width=7.5cm]{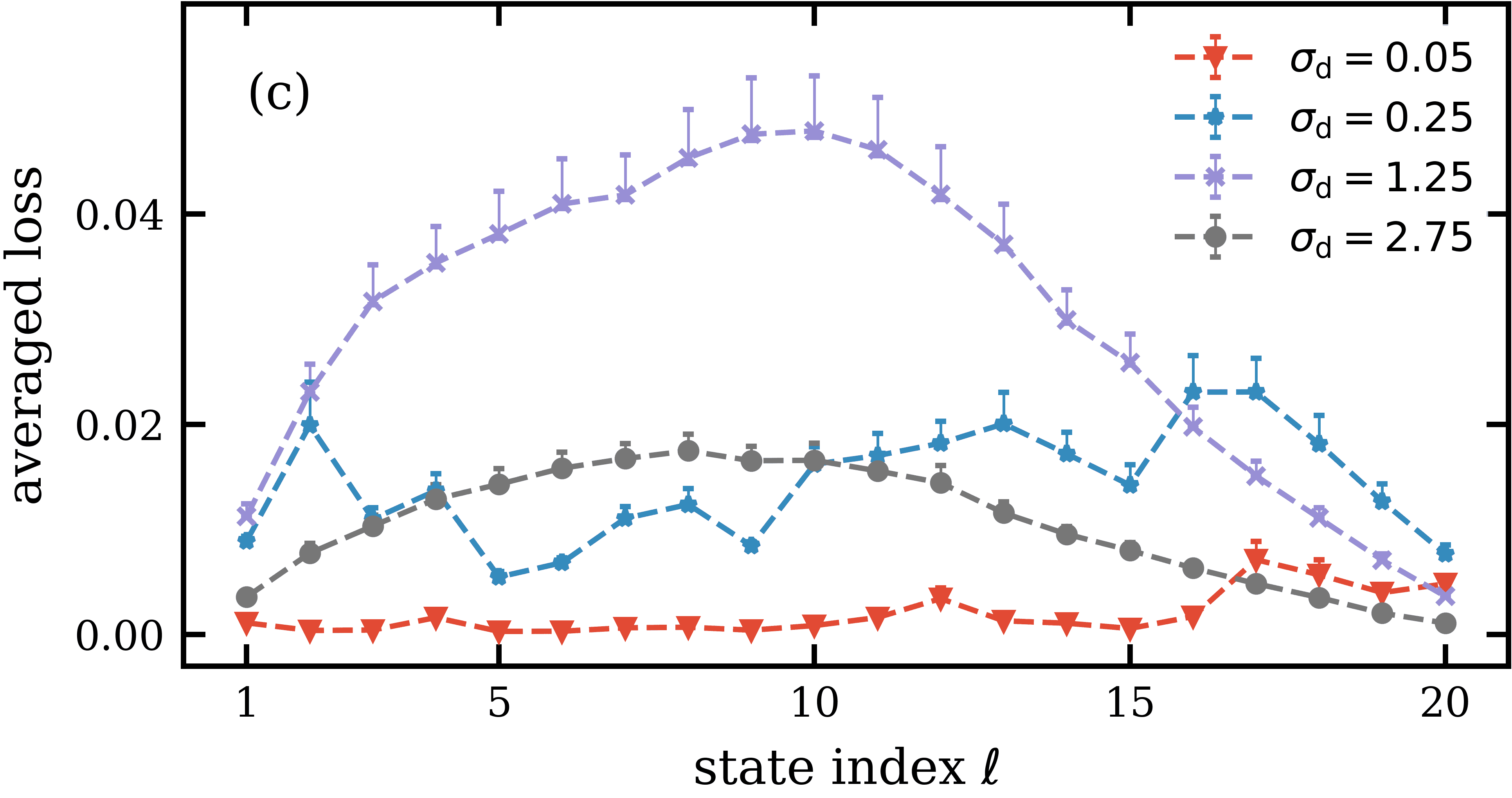}
  \end{minipage}
\caption{(a) Architecture of the neural network with the output shape of each layer shown in brackets. There are $417\,364$ trainable parameters in the neural network. (b) Training loss and validation loss during training. (c) Same as Fig.~\ref{fig:Architecture2D}(c) but now for the 1D case (the average is over $10\,000$ realizations)}
\label{fig:Architecture}
\end{figure*}

Training and validation followed a similar procedure as described in the 2D case. The architecture of the CNN is illustrated in Fig.~\ref{fig:Architecture} (a). There are $417\,364$ trainable parameters. As described in the Letter, the input data are 1D arrays containing $N_{\rm dip}$  entries (in the present case $N_{\rm dip}=512$). The outputs of the CNN are the predicted wave function coefficients which are arrays consisting of $N=20$ elements (for the considered case of an aggregate with 20 molecules). 
 
In order to train and validate the neural network, we prepare a huge  number of spectra stemming from a broad range of wave function coefficients. As described in the Letter we add disorder to the Hamiltonian Eq.~(1) in the Letter. We choose as disorder strengths $\delta \varepsilon_m \sim \mathcal N (0, \sigma_{\rm d}^2)$ for the molecular site energy $\varepsilon_m$ with the values  $\sigma_{\rm d} \in \{0.02,\ 0.04,\ 0.06,\ 0.08,\ 0.1,\ 0.2,\ 0.3,\ 0.4,\ 0.5,\ \\ 1.0,\ 1.5,\ \dots, 5.0\}$, which capture cases from weak localization to strong localization. In addition, we also generate some data by adding disorder to the transition dipole-dipole interaction, by using  $\tilde{V}_{mn}=\big[ \vec{\mu}_m\cdot \vec{\mu}_n - 3 (\vec{\mu}_m \cdot \frac{\vec{R}_{mn}}{R_{mn}})(\vec{\mu}_n \cdot \frac{\vec{R}_{mn}}{R_{mn}} ) + \delta V_{mn} \big] / R_{mn}^3$ with $\delta V_{mn} \sim \mathcal N (0, \sigma_{\rm od}^2)$ and $\sigma_{\rm od} \in [0.1, 0.2, 0.3, 0.4, 0.5]$. For each disorder strength, 10\,000 realizations are calculated, producing 4.6 million data samples in total. Randomly selected 80\% of these data are used for training and the rest for validation. 
 
The model is trained for 500 epochs with a batch size of 512. As shown in Fig.~\ref{fig:Architecture} (b), both the training loss and validation loss reach a small value around 0.012.

The testing data are obtained using the same procedure as that for the training data with disorder strengths $\sigma_{\rm d} = 0.05, 0.25, 1.25$, and $2.75$. For each disorder strength, $10\,000$ realizations are launched and the distribution of the loss-values are presented in the top row of Fig.~3 in the letter. The averaged loss for each is are shown in Fig.~\ref{fig:Architecture} (c).

\clearpage
\newpage
\subsection{\label{sec:Prediction_examples}Wave functions reconstructed with the trained CNN}

\begin{figure*}[bhtp]
\centering
\includegraphics[width=14.5cm]{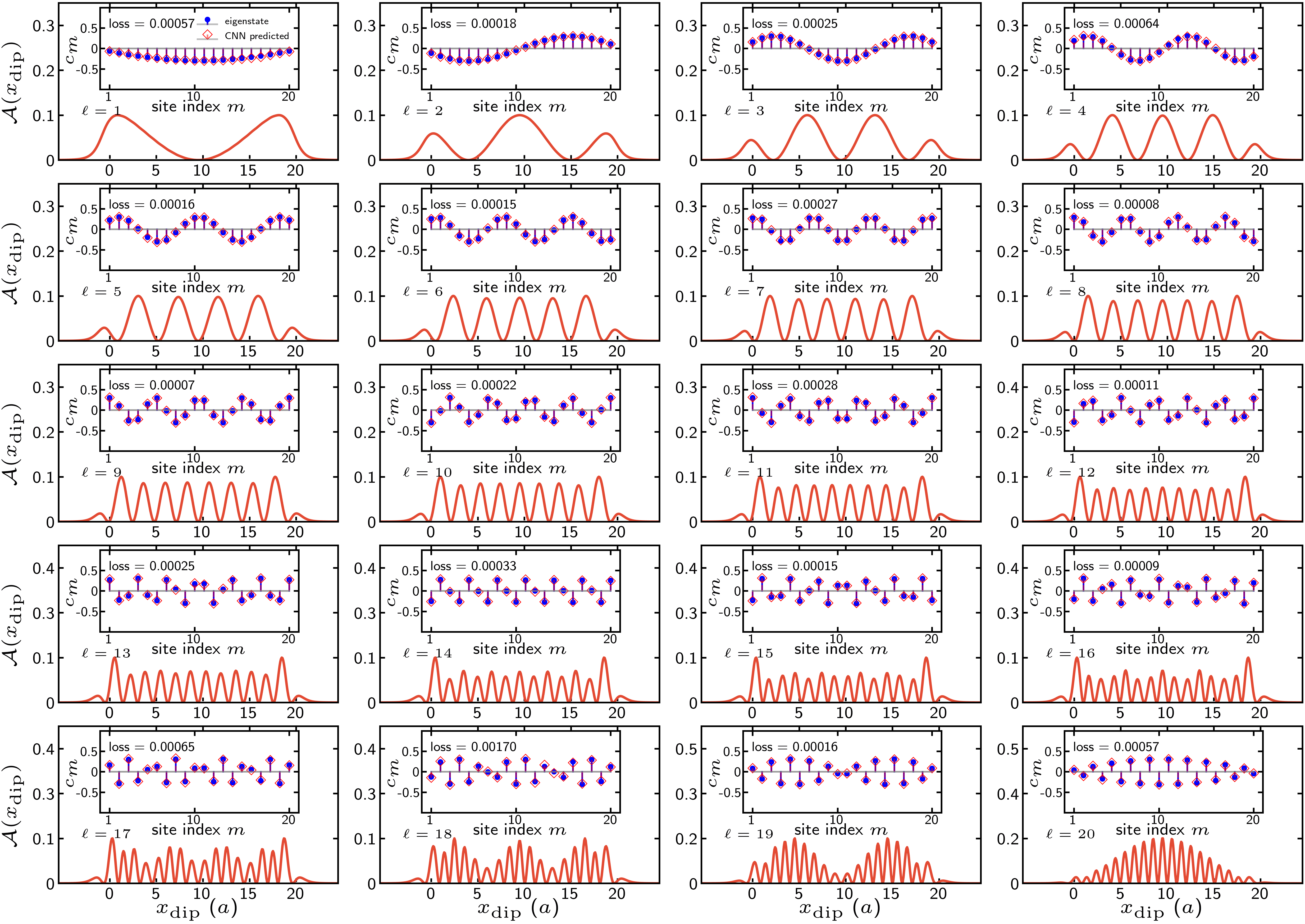}
\caption{Near-field spectra for the disorder-free 1D chain of 20 molecules. Comparison of the exact (blue dots) and predicted (red squares) wave function coefficients are shown in inserts.}
\label{fig:Test_NoDiorder}
\end{figure*}

For the case without disorder, the predicted wave functions for all the 20 eigenstate are displayed in Fig.~\ref{fig:Test_NoDiorder} together with the exact ones. The predicted  wave functions agree perfectly with the exact coefficients for all states. The loss values are as small as $10^{-4}$ for most of the states. 

\begin{figure*}[htp]
\centering
\includegraphics[width=14.5cm]{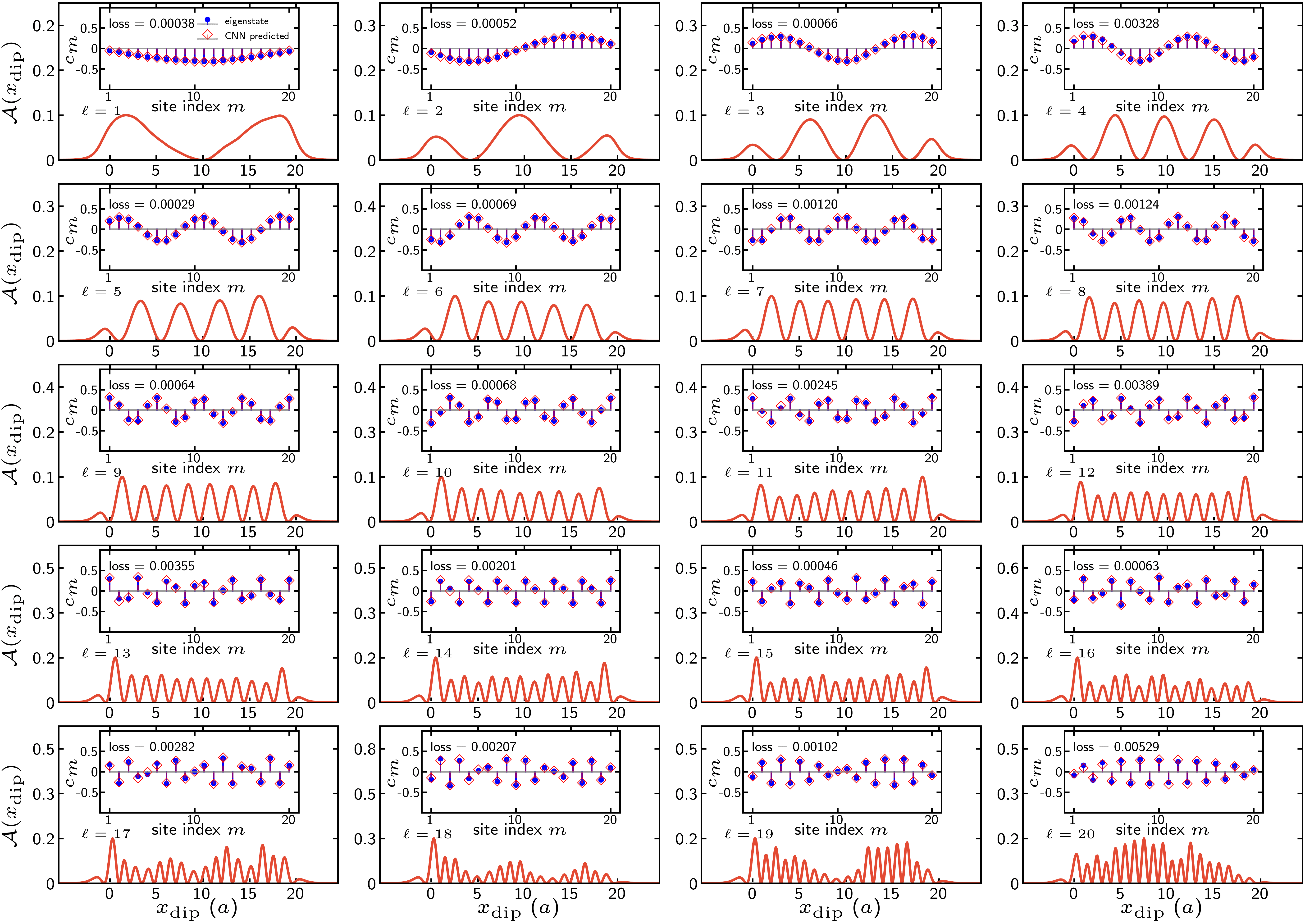}
\caption{Near-field spectra for a realization with diagonal disorder $\sigma_{\rm d} = 0.05$. Comparison of the exact (blue dots) and predicted (red squares) wave function coefficients are shown in inserts.}
\label{fig:Test_SmallDiagonal}
\end{figure*}

\begin{figure*}[htp]
\centering
\includegraphics[width=14.5cm]{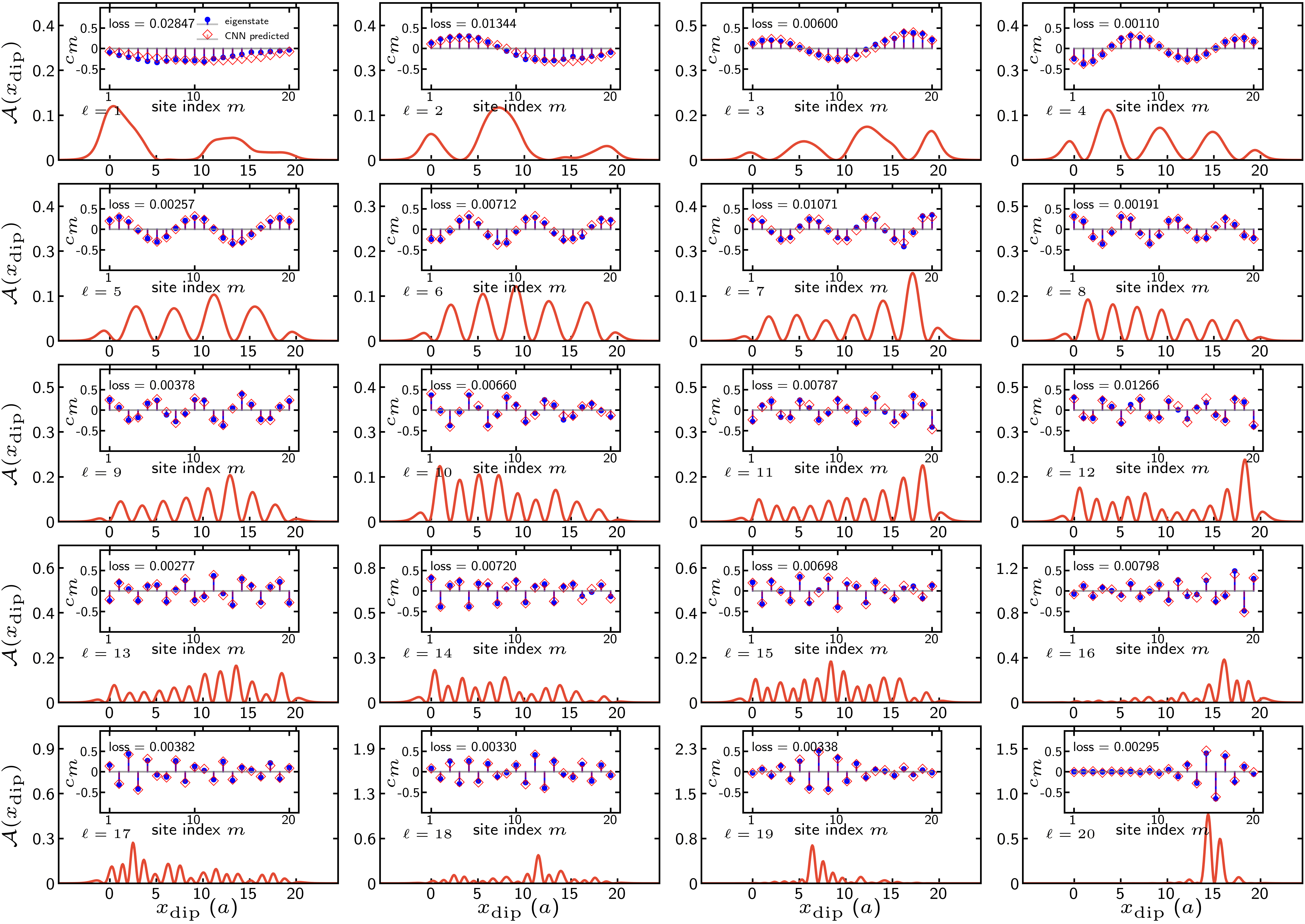}
\caption{As Fig.~\ref{fig:Test_SmallDiagonal} but for $\sigma_{\rm d} = 0.25$.
}
\label{fig:Test_IntermediateDiagonal}
\end{figure*}

\begin{figure*}[htp]
\centering
\includegraphics[width=14.5cm]{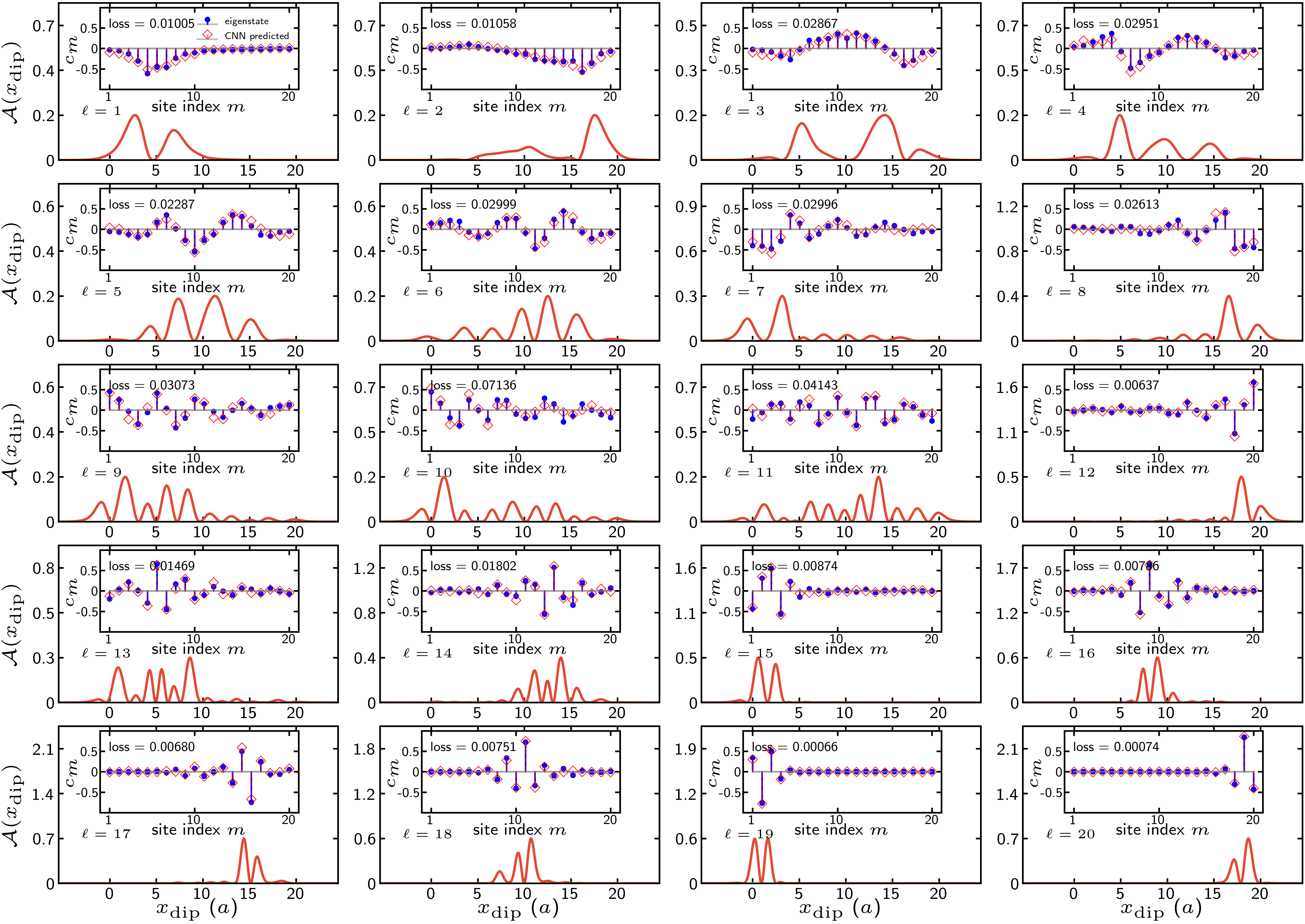}
\caption{As Fig.~\ref{fig:Test_SmallDiagonal} but for $\sigma_{\rm d} = 1.25$.
}
\label{fig:Test_LagerDiagonal_1}
\end{figure*}

\begin{figure*}[htp]
\centering
\includegraphics[width=14.5cm]{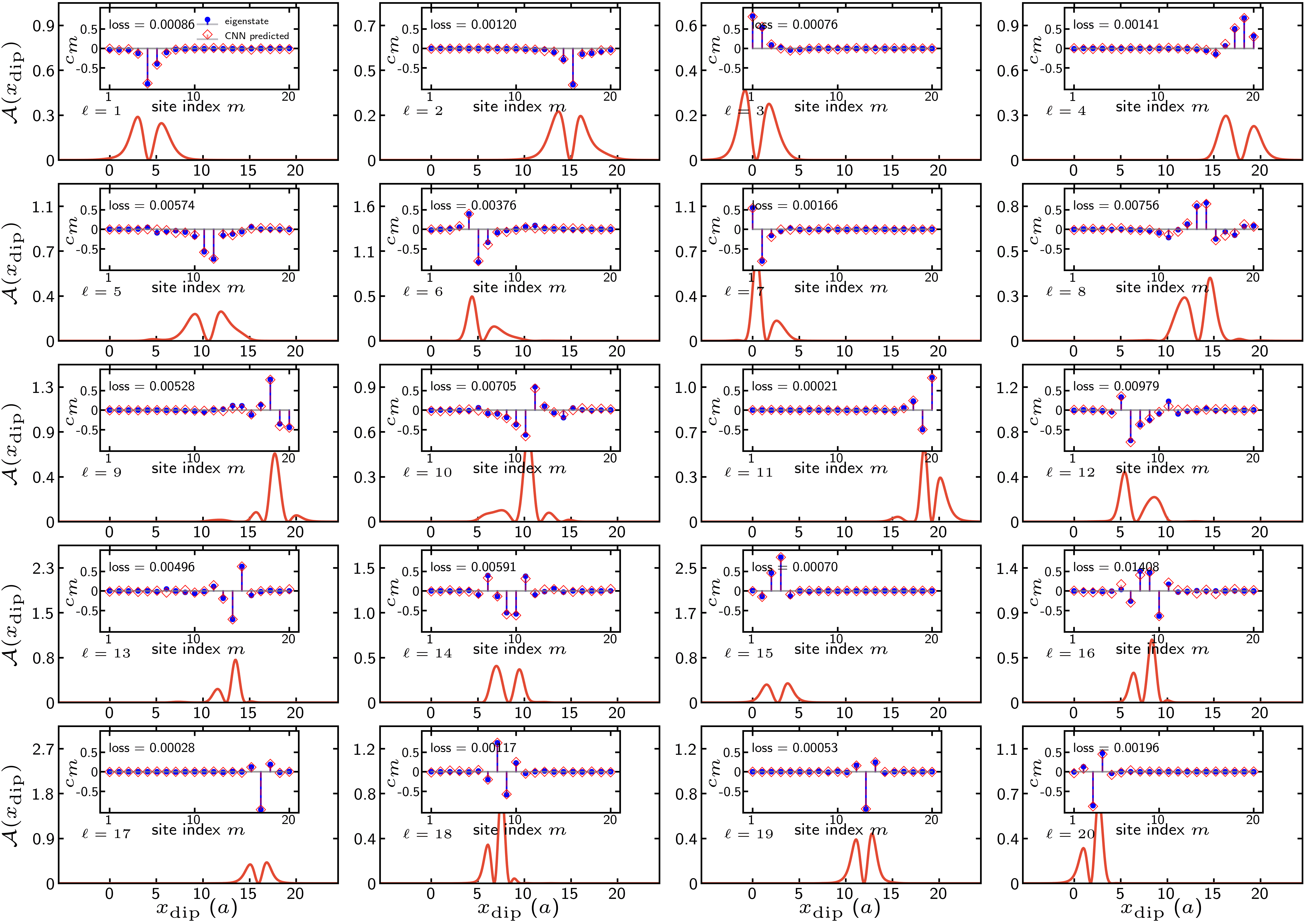}
\caption{As Fig.~\ref{fig:Test_SmallDiagonal} but for $\sigma_{\rm d} = 2.75$.
}
\label{fig:Test_LargeDiagonal}
\end{figure*}

\begin{figure*}[htp]
\centering
\includegraphics[width=14.5cm]{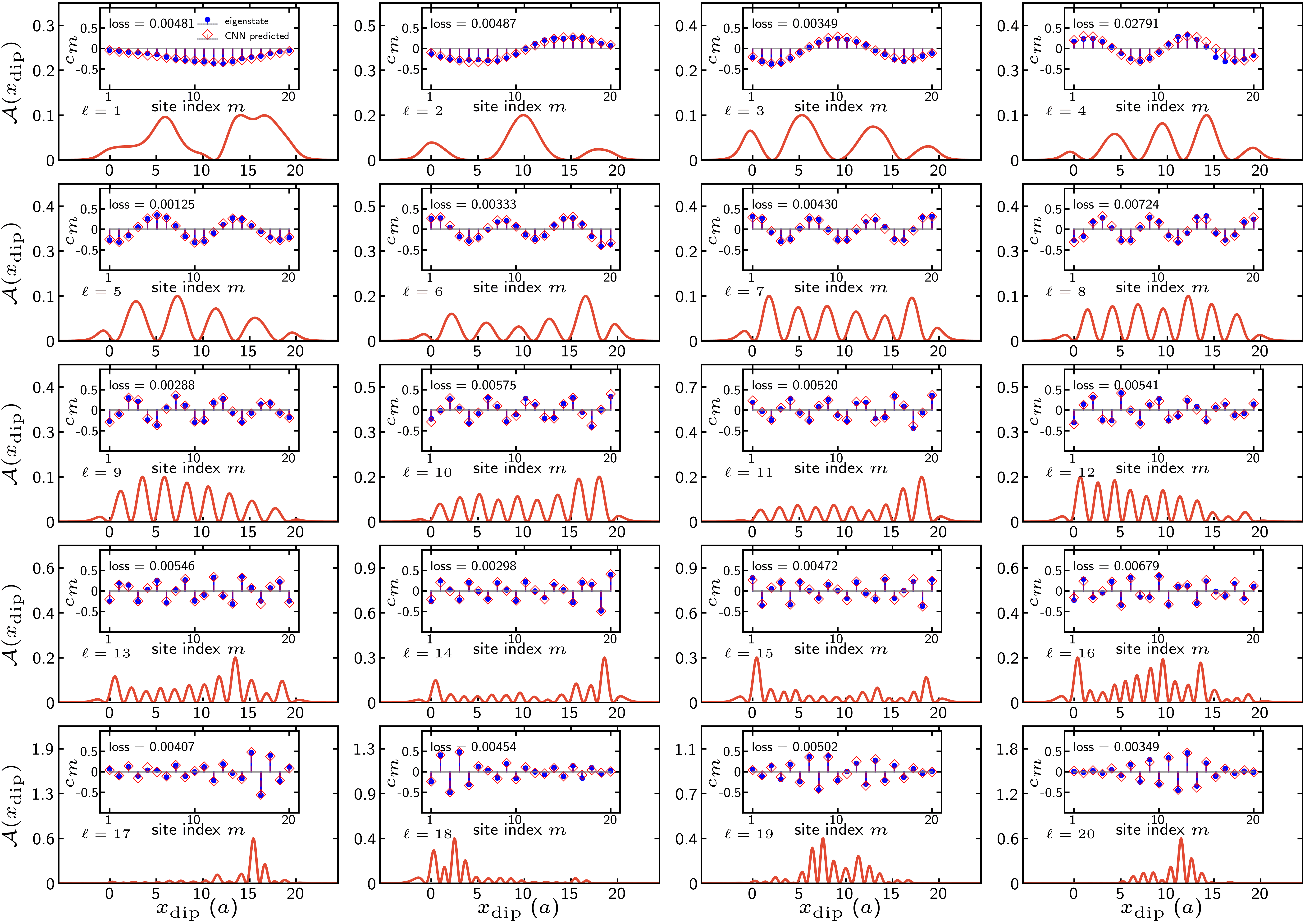}
\caption{As Fig.~\ref{fig:Test_SmallDiagonal} but for off-diagonal disorder $\sigma_{\rm od} = 0.25$.
}
\label{fig:Test_offdiagonal}
\end{figure*}

For the disorder case, we tested $\sigma_{\rm d} = 0.05$, $0.25$, $1.25$, and $2.75$ for the disorder added to the site energies, and $\sigma_{\rm od} = 0.25$ for the disorder added to the inter-molecular coupling. To obtain a feeling how typical wave function look for the chosen disorder, in Figs.~\ref{fig:Test_SmallDiagonal}-\ref{fig:Test_offdiagonal} we present results for one single realization for each disorder strength and plot the comparisons between the exact coefficients and the predicted ones. 

\clearpage
\newpage
\subsection{\label{sec:low_res} Wave function reconstruction using the CNN trained with lower resolution spectra}

In the 1D cases discussed above, we have evaluated the spectra at $512$ spatial positions $x_{\rm dip}$ of the tip. The total range of the sampled $x_{\rm dip}$-values was $40\, a$, symmetrically around the aggregate (which has a length of $20\, a$). That means we have a spatial resolution of about $13$ data points for the center to center distance $a$ between neighboring molecules. For a center to center distance $a\approx 1.25\,\mathrm{nm}$ this corresponds to a resolution of about $1\,$ \AA.
 
\begin{figure*}[htp]
\centering
\includegraphics[width=4.5cm]{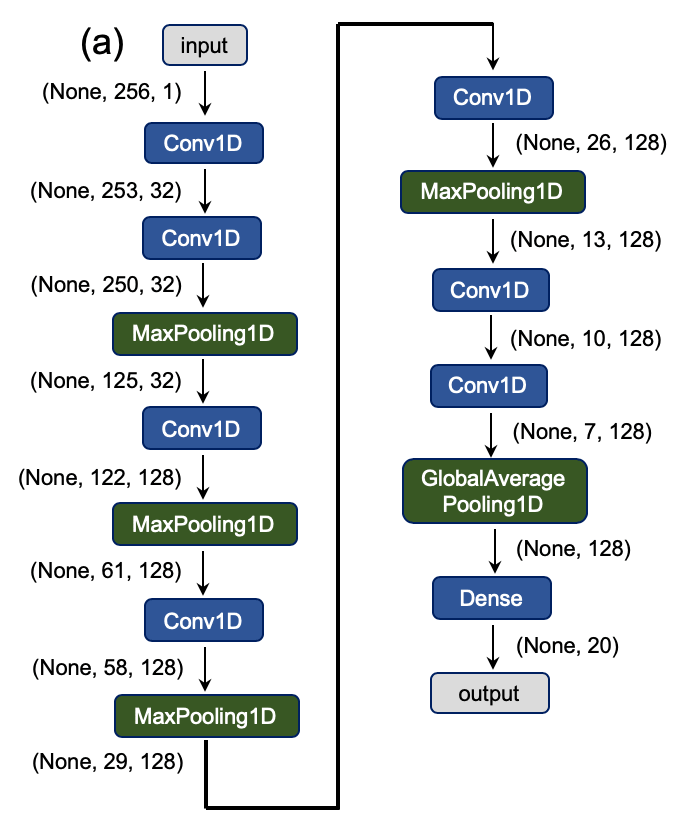}
\hspace{2cm}
\includegraphics[width=8.cm]{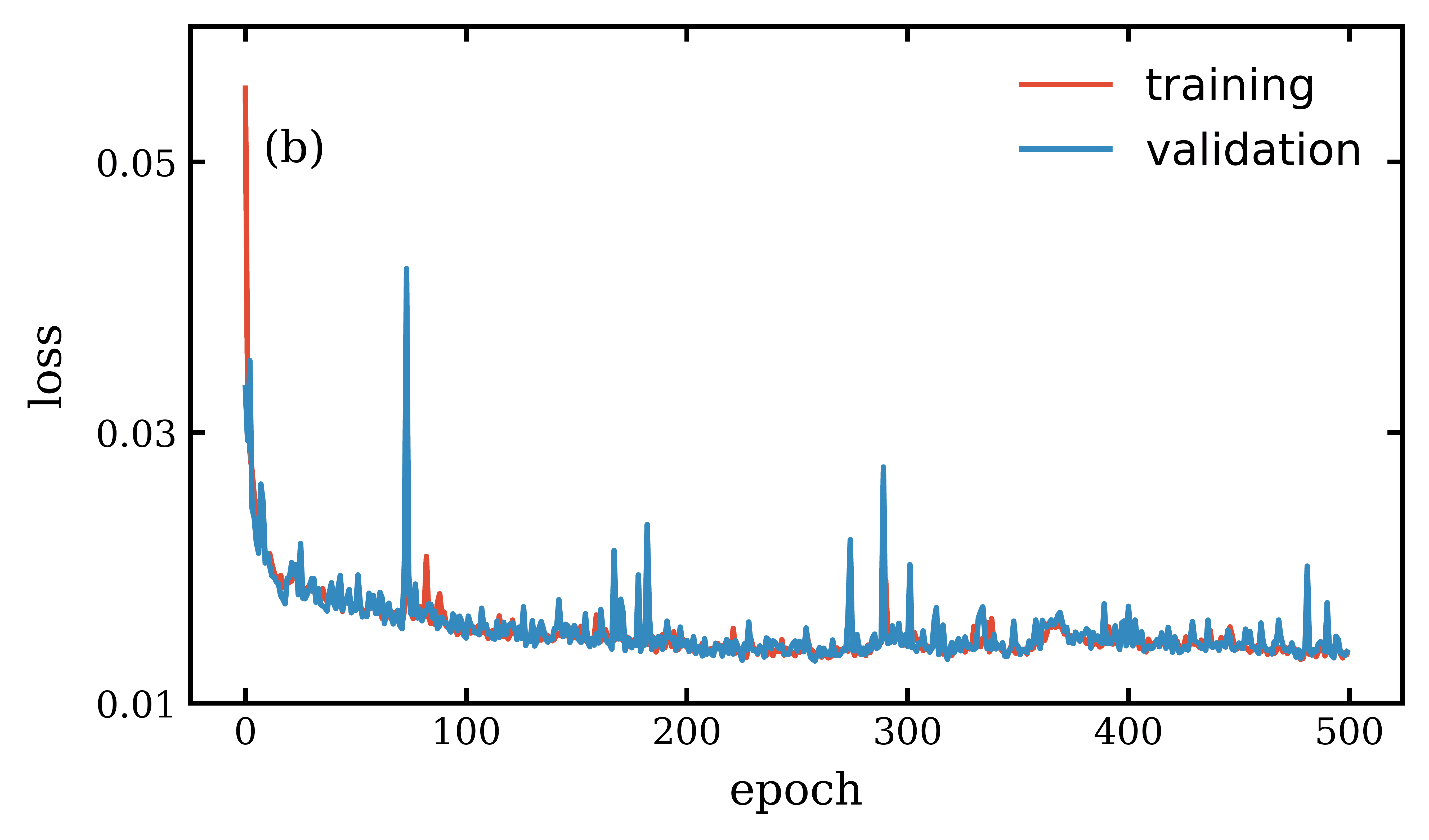}
\includegraphics[width=14.5cm]{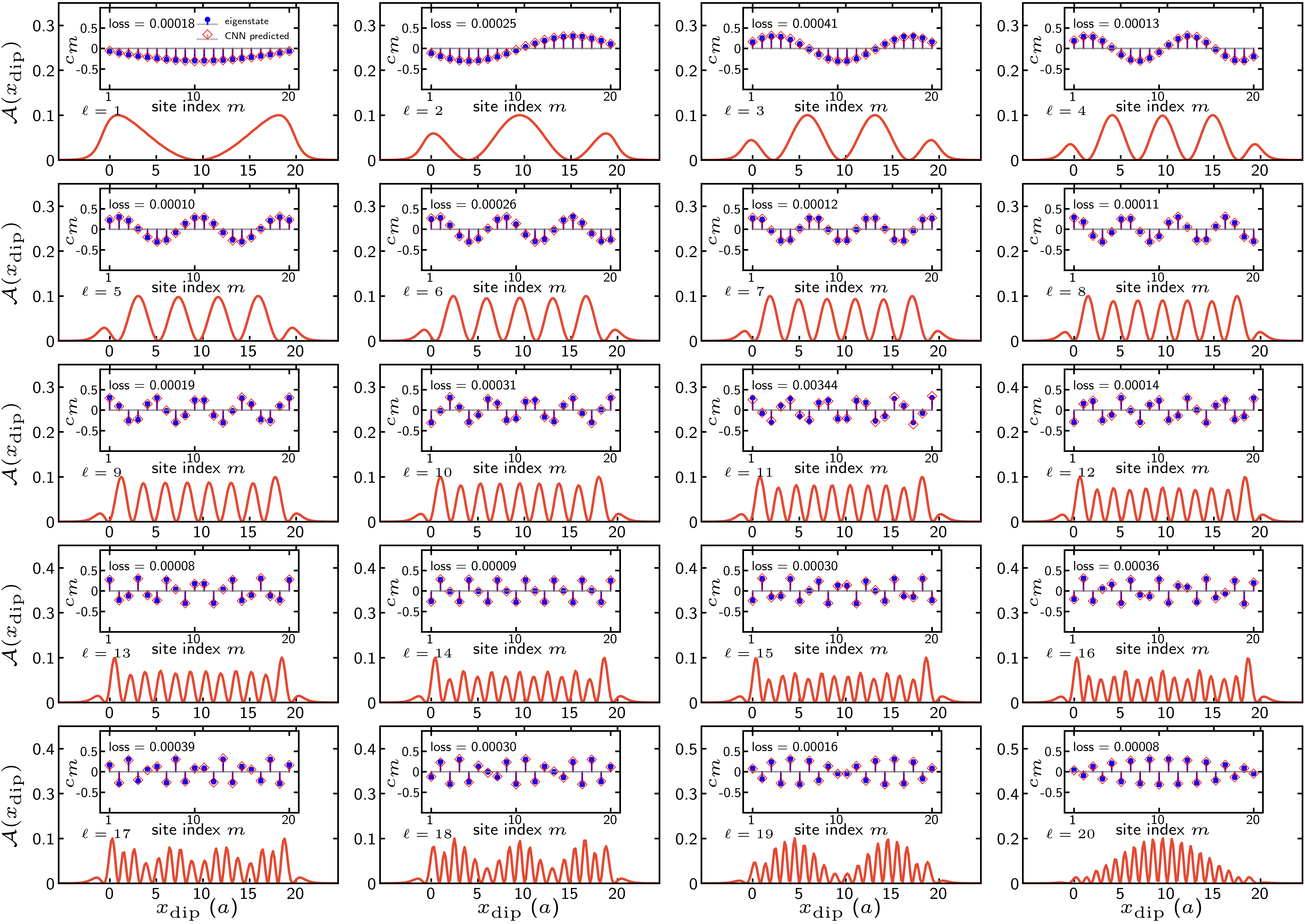}
\caption{Wave function coefficients reconstruction using CNN trained with spectra evaluated at $256$ tip positions. The architecture is shown in (a), and training and validation loss are plotted (b). Presented in bottom panel are the $256$-points spectra for the disorder-free chain. Comparison of the exact (blue dots) and predicted (red squares) wave function coefficients are shown in inserts.}
\label{fig:spectra256}
\end{figure*}

\begin{figure*}[htp]
\centering
\includegraphics[width=5cm]{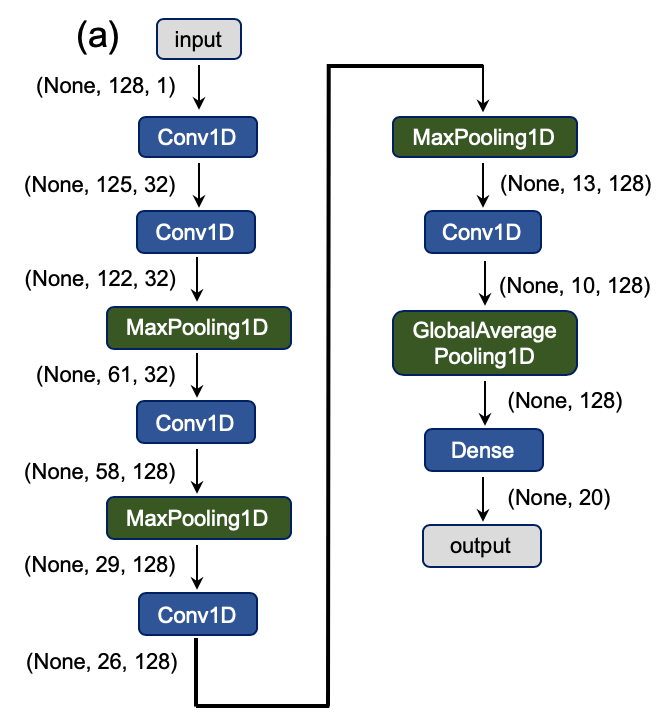}
\hspace{1.5cm}
\includegraphics[width=8cm]{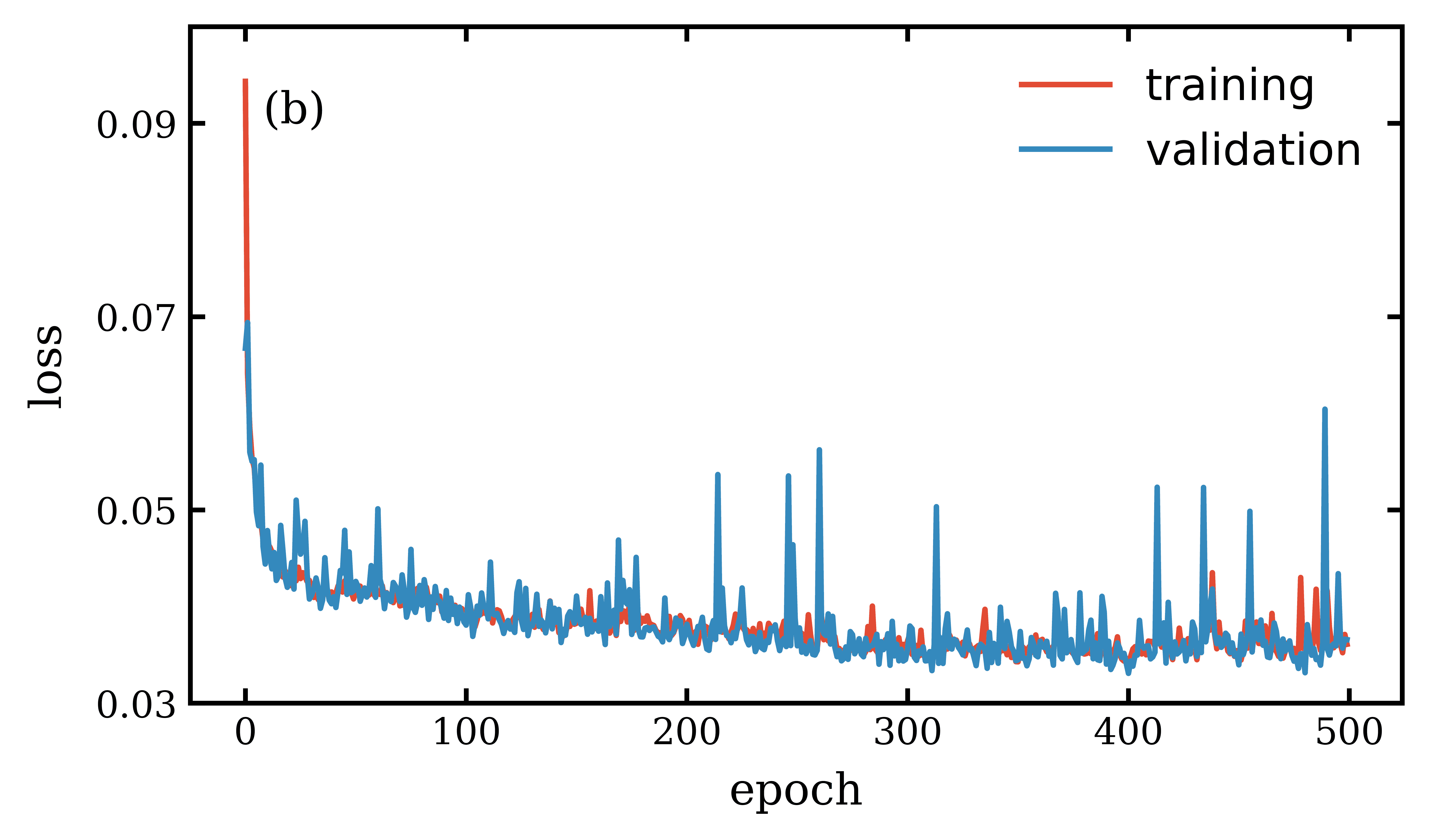}
\includegraphics[width=14.5cm]{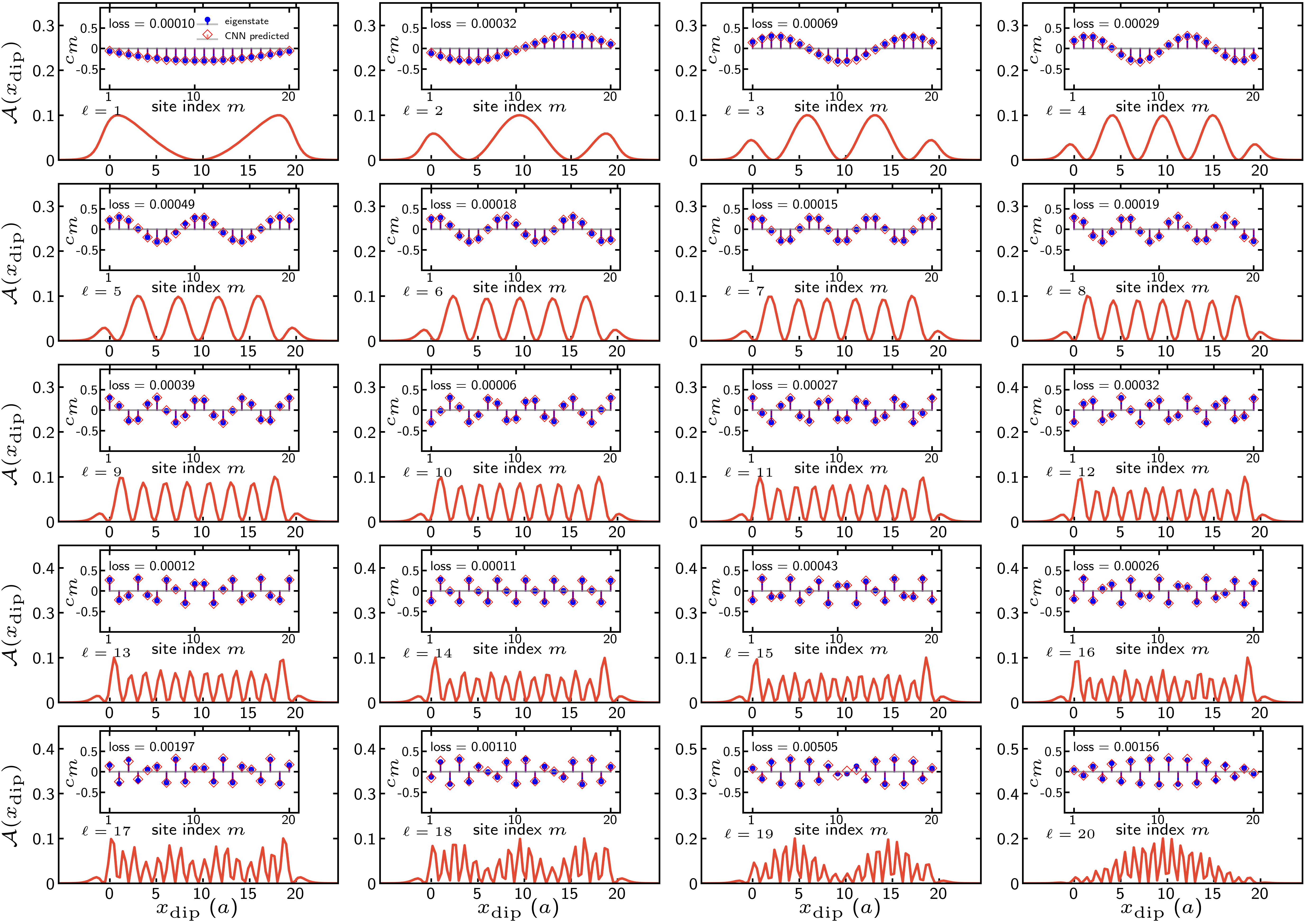}
\caption{As Fig.~\ref{fig:spectra256} but for spectra evaluated at $128$ tip positions. 
}
\label{fig:spectra128}
\end{figure*}

To explore the validity of the architecture to the spatial resolution of the spectra, we also train the CNN using the spectra with reduced resolutions. Because of the resulting reduction of the size of the input, we have also changed the architecture. The new architectures and the training are shown in Fig.~\ref{fig:spectra256} and \ref{fig:spectra128} for the case of $N_{\rm tip}=256$ and  $N_{\rm tip}=128$, respectively.

In Figs.~\ref{fig:spectra256}  and \ref{fig:spectra128} we also show comparisons between the exact eigenstate coefficients for the disorder-free Hamiltonian and the predict values from the CNN. For both cases, the trained CNN can perfectly reconstruct all the eigenstate coefficients, indicating that the wave function reconstruction from near-field spectra is also feasible by using the CNN trained with low resolution spectra. One finds that the averaged loss slightly grows upon decreasing the resolution ($3.7\cdot 10^{-3}$ for the case of 256 tip positions and $7.0\cdot 10^{-3}$ for 128 tip positions).

\clearpage
\newpage

\subsection{\label{sec:noise} Noise added to the  spectra}
Here we provide more information about the case when noise is added to the spectrum.

\begin{figure}[h]
\includegraphics[width=8.cm]{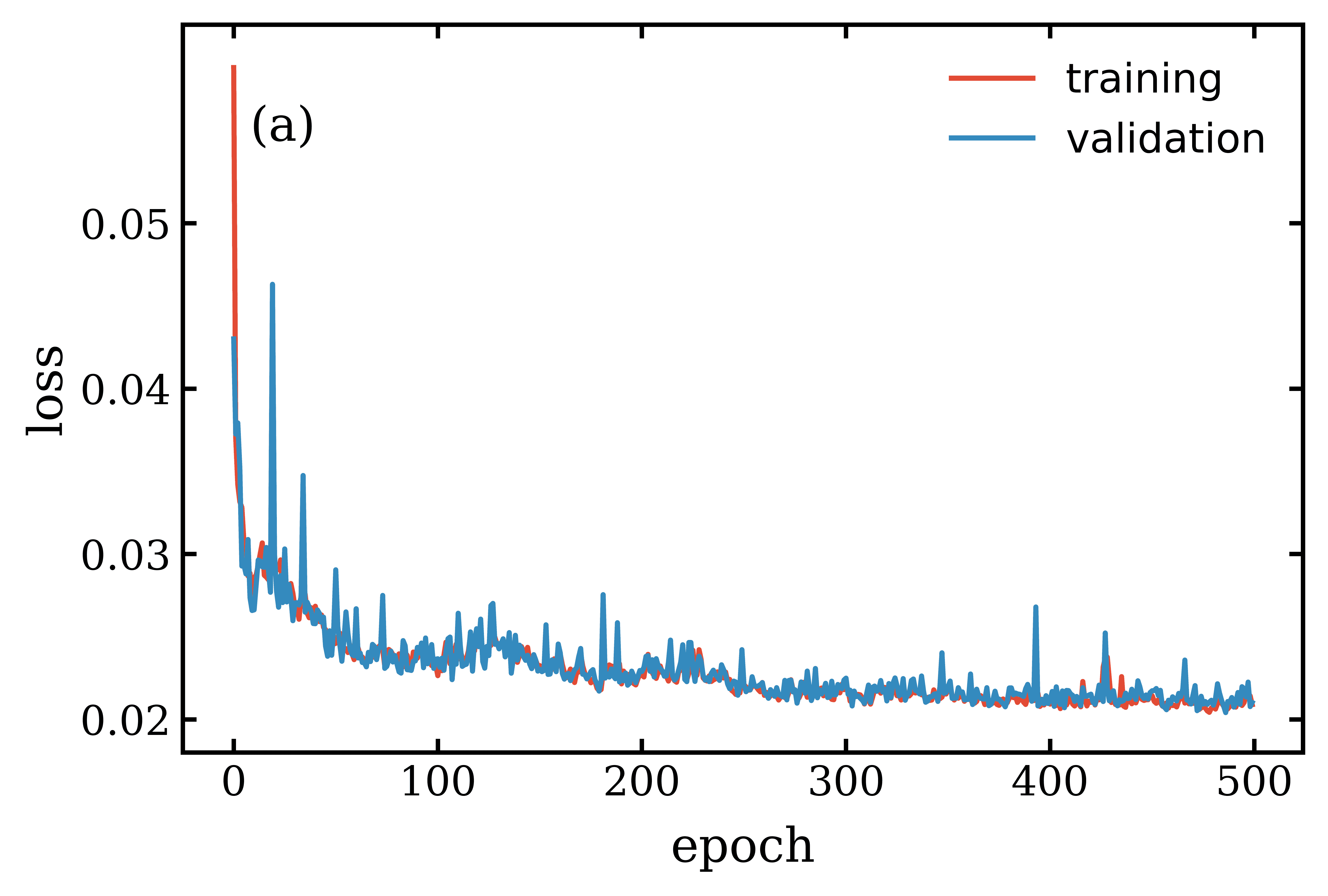}
\hspace{1cm}
\includegraphics[width=8.cm]{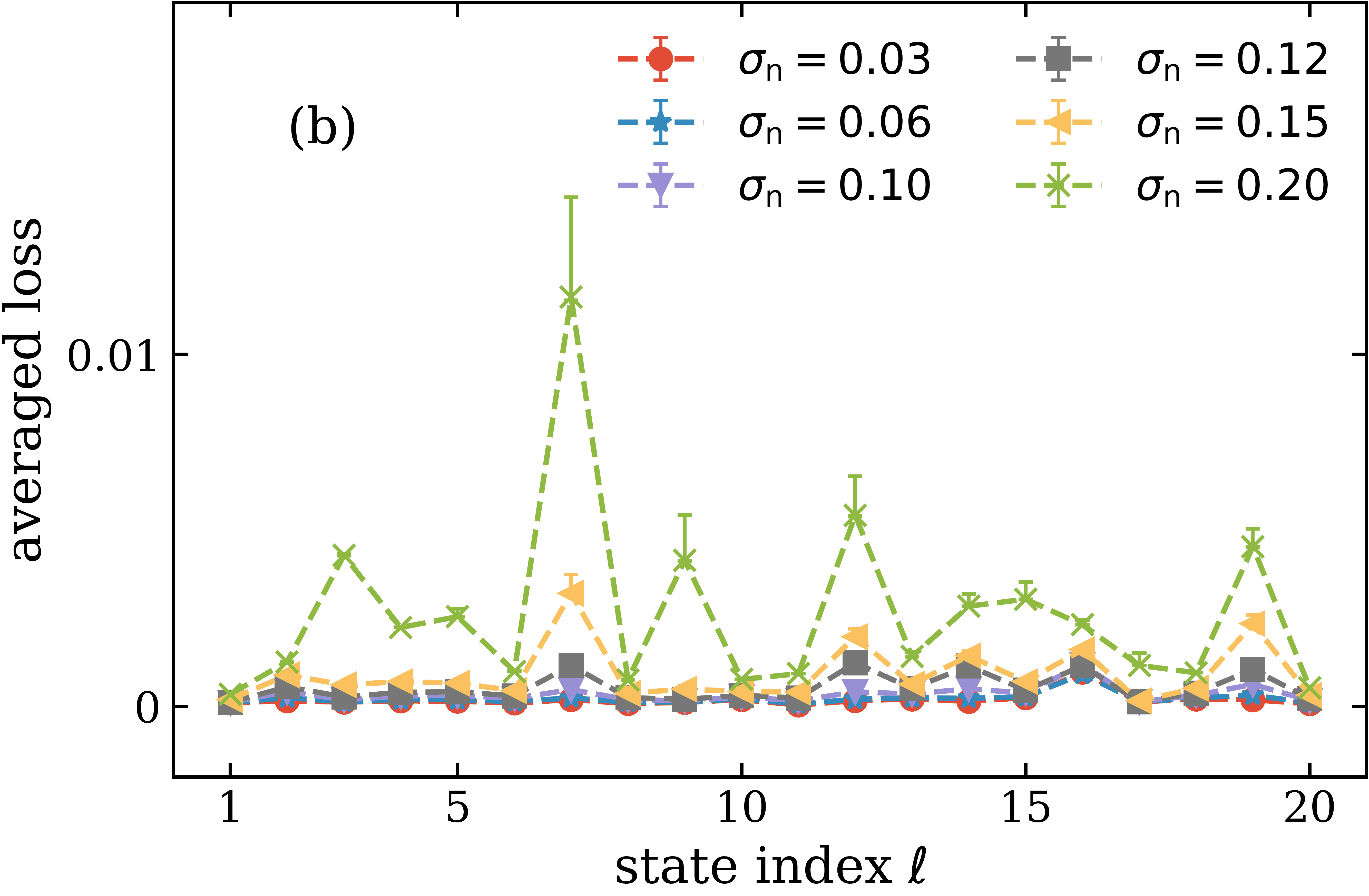}
\includegraphics[width=17cm]{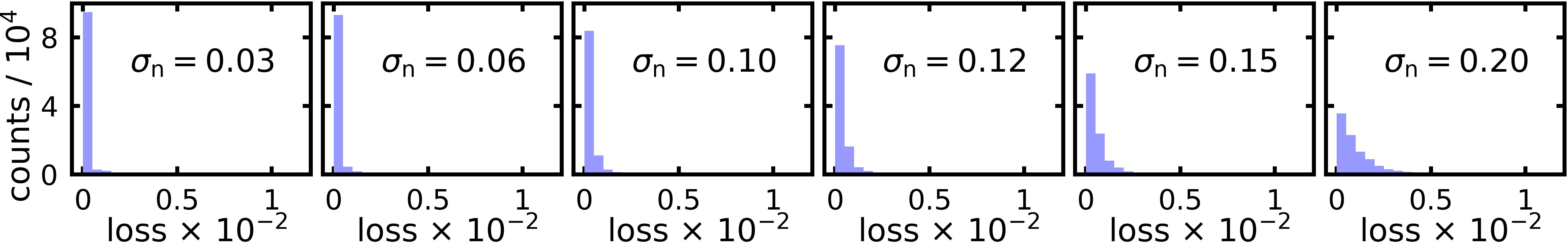}
\caption{\label{fig:nois_training} (a) Training of the CNN using spectra where noise with $\sigma_\mathrm{n}=0.1$ is added. The network is trained for 500 epochs with a batch size 512. (b) averaged loss value per state (average over 5000 realizations). Bottom row: As Fig.~4 of the Letter, but for additional values of $\sigma_{\rm n}$.}
\end{figure}

\begin{figure}[hp]
\includegraphics[width=14.5cm]{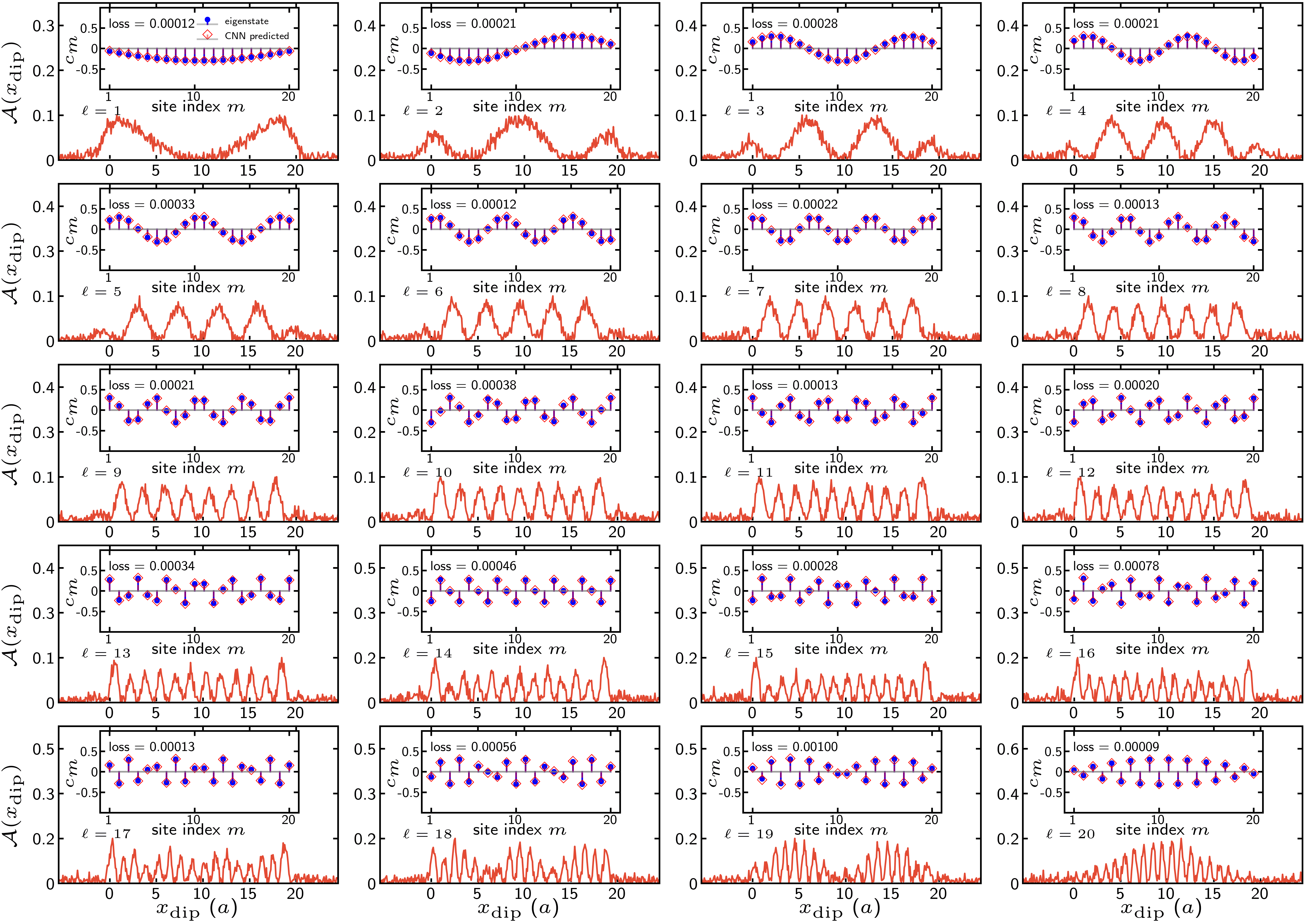}
\caption{\label{fig:noise_0.1}Spectra of the ideal linear chain (without disorder) but with noise added. All eigenstates are shown. To each spectrum a realization of the noise with $\sigma_\mathrm{n}=0.1$ is added.
Note, that the CNN is trained with this noise strength. The exact (blue dots) and predicted (red squares) wave function coefficients are shown in inserts.}
\end{figure}

\begin{figure}[hp]
\includegraphics[width=14.5cm]{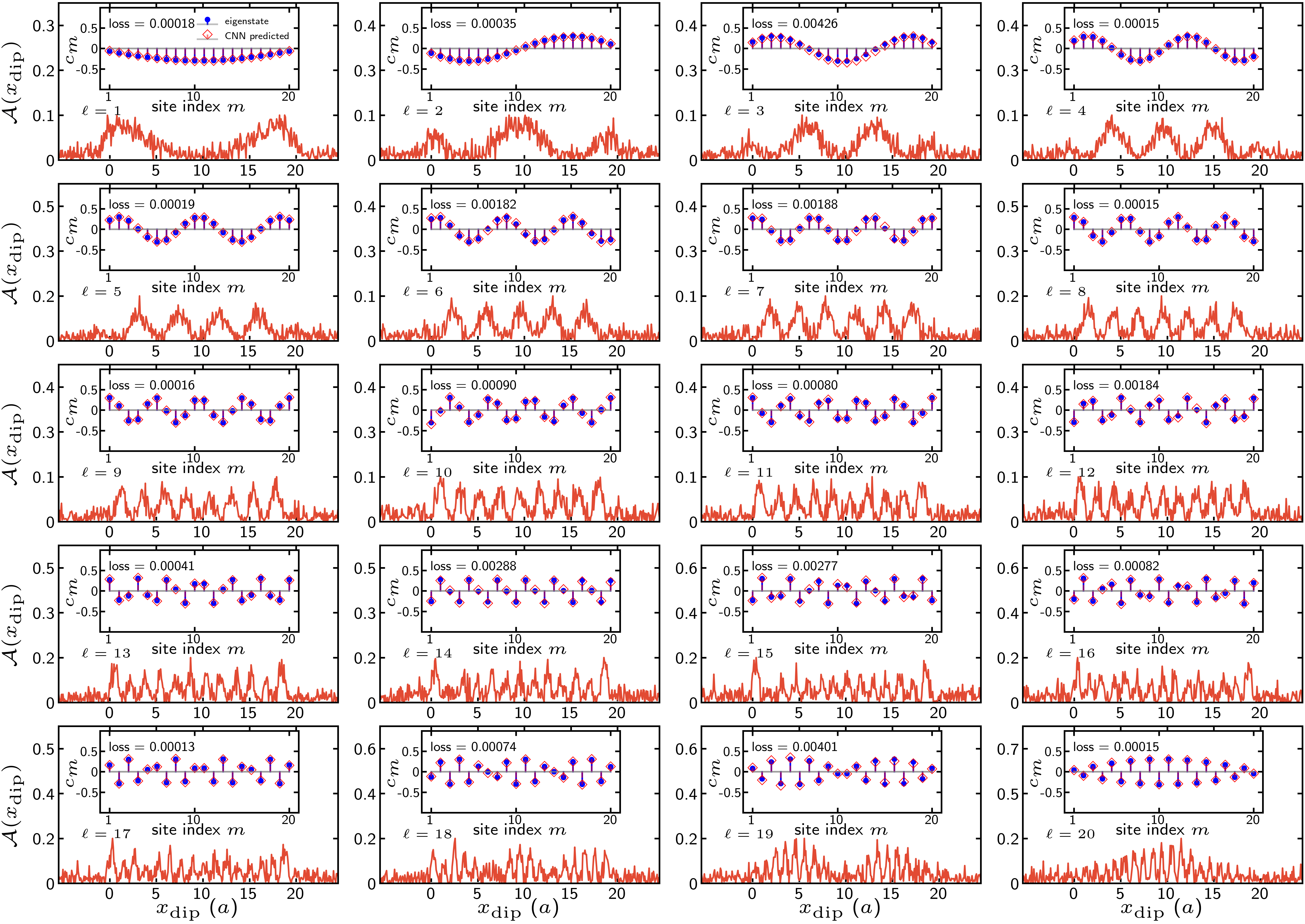}
\caption{\label{fig:noise_0.2}As Fig.~\ref{fig:noise_0.1} but for $\sigma_\mathrm{n}=0.2$.}
\end{figure}

The CNN has the same architecture as the one use for the 1D case without noise (the spectra are evaluated at 512 tip positions). As described in the Letter, we have trained the CNN using the same 4.6 million realizations as in the cases considered above, but now to each spectrum we add a random realization of the noise with relative strength  $\sigma_\mathrm{n}=0.1$. Examples of such spectra used for training are given in Fig.~\ref{fig:noise_0.1}. In the same figure also the corresponding predicted wave functions and the exact wave functions are shown. In Fig.~\ref{fig:noise_0.2} spectra with disorder strength $\sigma_\mathrm{n}=0.2$ together with the respective predicted wave functions are presented. The training and validation loss for the CNN trained with noisy spectra are displayed in Fig.~\ref{fig:nois_training} (a).

5000 realizations are performed for each scaling factor for the spectra of disorder-free system, and the distribution of the loss is used to plot the top row of Figure 4 in the Letter. Loss distributions for more scaling factors are shown in the bottom row of Fig.~\ref{fig:nois_training} and the averaged loss for all states with different noise levels is illustrated in Fig.~\ref{fig:nois_training} (b).


\clearpage
\newpage

\subsection{\label{sec:polarizability}Homogeneous line-broadening and tip-effects}

To include homogeneous broadening and the interaction with the tip we have used the polarizability based formalism used in \cite{Gao2018_SI}. We use also the same material parameters as in that reference. This formalism reduces in the limit $\gamma \rightarrow 0$ and tip-diameter $\rightarrow 0$ to the model described by Eqs.~(1)-(3) of the Letter.

In the following we provide the relevant formulas and the parameters used in our calculations. The basic quantity to be determined are the dipole moments $\vec{P}_{m}(\omega)$ of the particle. We use the following convention to label the particles: The index 0 stands for the tip (which we model as polarizable sphere) and the indices 1 to $N$ refer to the molecules (as in the Letter). The dipole moment of particle $m$ is induced by the total field at this particle, which consists of the external field $\vec{E}_{m}^{\rm ext} (\omega) = \vec{E} (\vec{R}_{m})$ and the ``internal'' fields $\vec{E}_{mn}^{\rm int}$ produced by all other particles $n$:
\begin{equation}
\vec{P}_{m} (\omega) = \overleftrightarrow{\alpha}_{m} (\omega) \Bigg( \vec{E}_{m}^{\rm ext} (\omega) + \sum_{n,n\neq m}^{N} \vec{E}_{mn}^{\rm int} \Bigg)
\end{equation}
where $\overleftrightarrow{\alpha}_{m} (\omega)$ is the frequency dependent polarizability. The external field is provided by a Hertzian dipole with dipole moment $\vec{d}$, located at $\vec{R}_{\rm dip}$ and is written as $\vec{E}_{m}^{\rm ext} =\vec{E}(\vec{R}_m;\vec{R}_{\rm dip})$ as given in Eq.~(\ref{eq:hertz_near}). The external field at the position of the tip is zero: $\vec{E}_{0}^{\rm ext} = 0$. The ``internal'' field is given by
\begin{equation}
\vec{E}_{mn}^{\rm int} (\omega) = - \overleftrightarrow{T}_{mn} \vec{P}_{n} (\omega)
\end{equation} 
with 
\begin{equation}
\overleftrightarrow{T}_{mn} = \frac{1}{4\pi\epsilon_{0}} \frac{1}{R_{mn}^3} \Bigg( \overleftrightarrow{I} - 3 \frac{\vec{R}_{mn} \otimes \vec{R}_{mn} } {R_{mn}^2} \Bigg)
\end{equation}
where $\otimes$ denotes the outer product, $\vec{R}_{mn}$ is the separation vector between $m$ and $n$, and $\overleftrightarrow{I}$ is the identity tensor. The absorption spectrum is obtained from 
\begin{equation}
\mathcal{A} (\omega) = {\rm Im} \sum_{m=0}^{N} \vec{P}_{m}(\omega) \cdot \vec{E}_{m}^{\rm ext} (\omega).
\end{equation}
The polarizability of the tip is modeled by the polarizability of a sphere with radius $a_{r}$
\begin{equation}
\label{eq:alpha_tip}
\overleftrightarrow{\alpha}_{0} (\omega) = \overleftrightarrow{\alpha}_{\rm tip} (\omega) = - 4 \pi \epsilon_{0} a_{r}^3 \frac{\epsilon({\omega}) - \epsilon_{\rm env}}{\epsilon({\omega}) + 2 \epsilon_{\rm env}} \overleftrightarrow{I},
\end{equation}
where $\epsilon_{\rm env}$ is the dielectric constant of the surrounding medium. The complex dielectric function $\epsilon({\omega})$ is evaluated according to a generalized Drude model with finite-size effects correction for particles with radius below $\sim 5$ nm,
\begin{equation}
\epsilon (\omega) = \epsilon_{\rm b} + \frac{\omega_{p}^2}{\omega(\omega - i \gamma_{p})} - \frac{\omega_{p}^2}{\omega(\omega - i \gamma_{p} - i \nu_{F} / a_{r})}.
\end{equation}
Here $\epsilon_{b}$ is an adjustable constant, and $\omega_{p}$, $\gamma_{p}$, and $\nu_{F}$ are the plasma frequency, Ohmic damping constant, and Fermi velocity in the bulk material of the tip, respectively. The polarizability tensor for the $m$th monomer is taken as
\begin{equation}
\overleftrightarrow{\alpha}_{m} (\omega) = - \frac{\vec{\mu}_{m} \otimes \vec{\mu}_{m}}{ \omega - \omega_{m} + i \gamma_{m}},
\end{equation}
where $\vec{\mu}_{m}$, $\omega_{m}$, and $\gamma_{m}$ are the transition dipole moment, transition frequency, and the damping constant for molecule $m$.

The parameters used in the calculations are listed in the table below
\begin{center}
  \begin{tabular}{ |c|c|c|c|c|c|c|c|c|c| } 
   \hline
   $\omega_{m}$ [ cm$^{-1}$] & $\gamma_{m}$ [ cm$^{-1}$] & $\mu_{m}$ [ Debye ] & $a$ [nm] & $a_{r}$ [nm] & $\epsilon_{b}$ & $\epsilon_{\rm env}$ &  $ \omega_{p} $ [ cm$^{-1}$ ] & $ \gamma_{p} $ [ cm$^{-1}$ ] & $\nu_{F}$ [ cm $s^{-1}$ ]\\ 
   \hline
   $2 \times 10^{4}$ & 1,\dots ,30 & 7.4 & 1.25 &  2.5 & 9 & 1  &  $7.26 \times 10^{4}$ & 400 & $ 1.39 \times 10^{8}$ \\ 
   \hline
   \end{tabular}
\end{center}

\newpage

In the following we investigate what predictions the CNN (trained as described in the Letter) makes if the spectra are subject to finite $\gamma_m$ and finite tip diameter.

\begin{figure}
  \centering 
   \includegraphics[width=6cm]{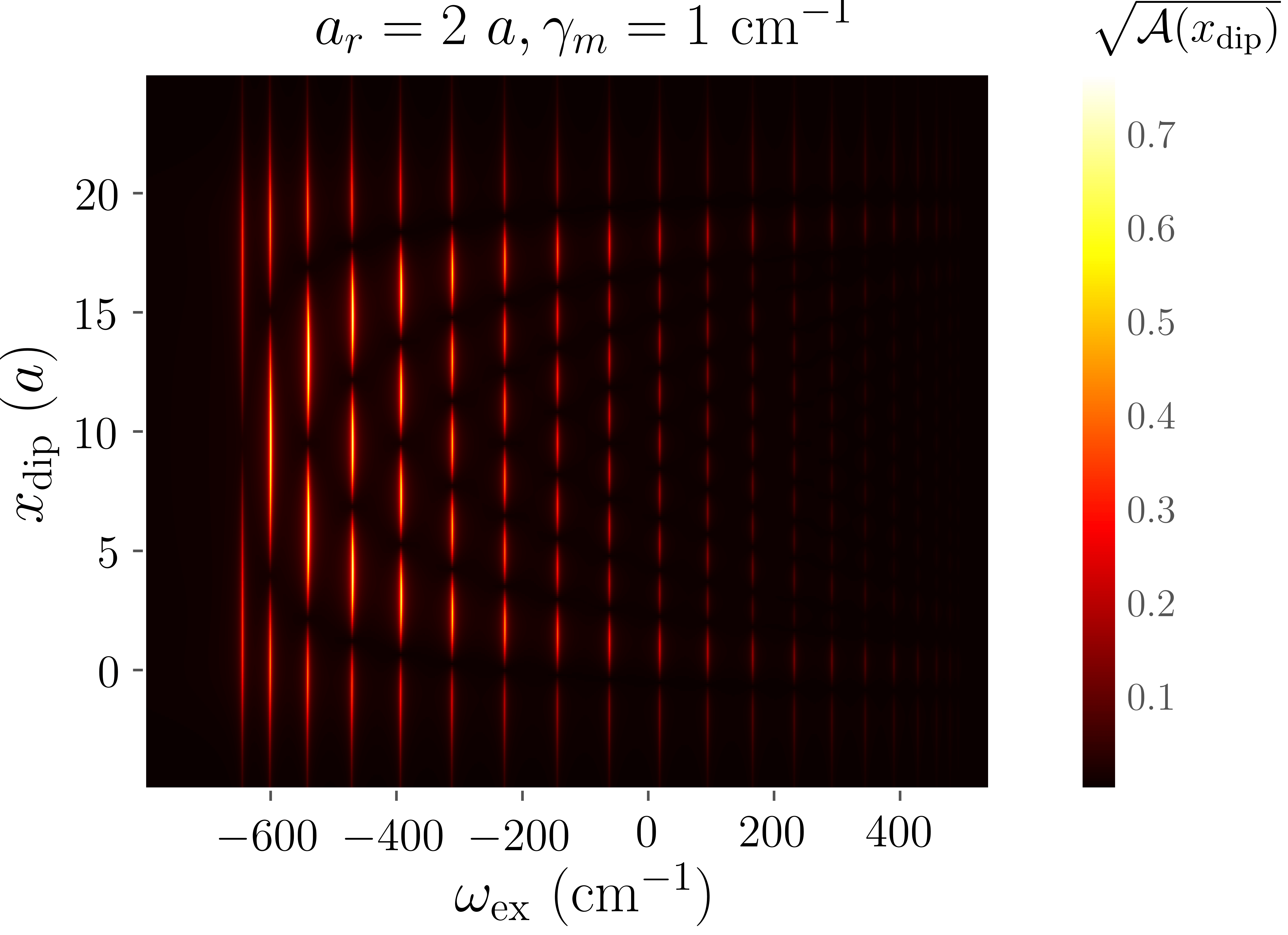}   
   \hspace{1cm}       
   \includegraphics[width=6cm]{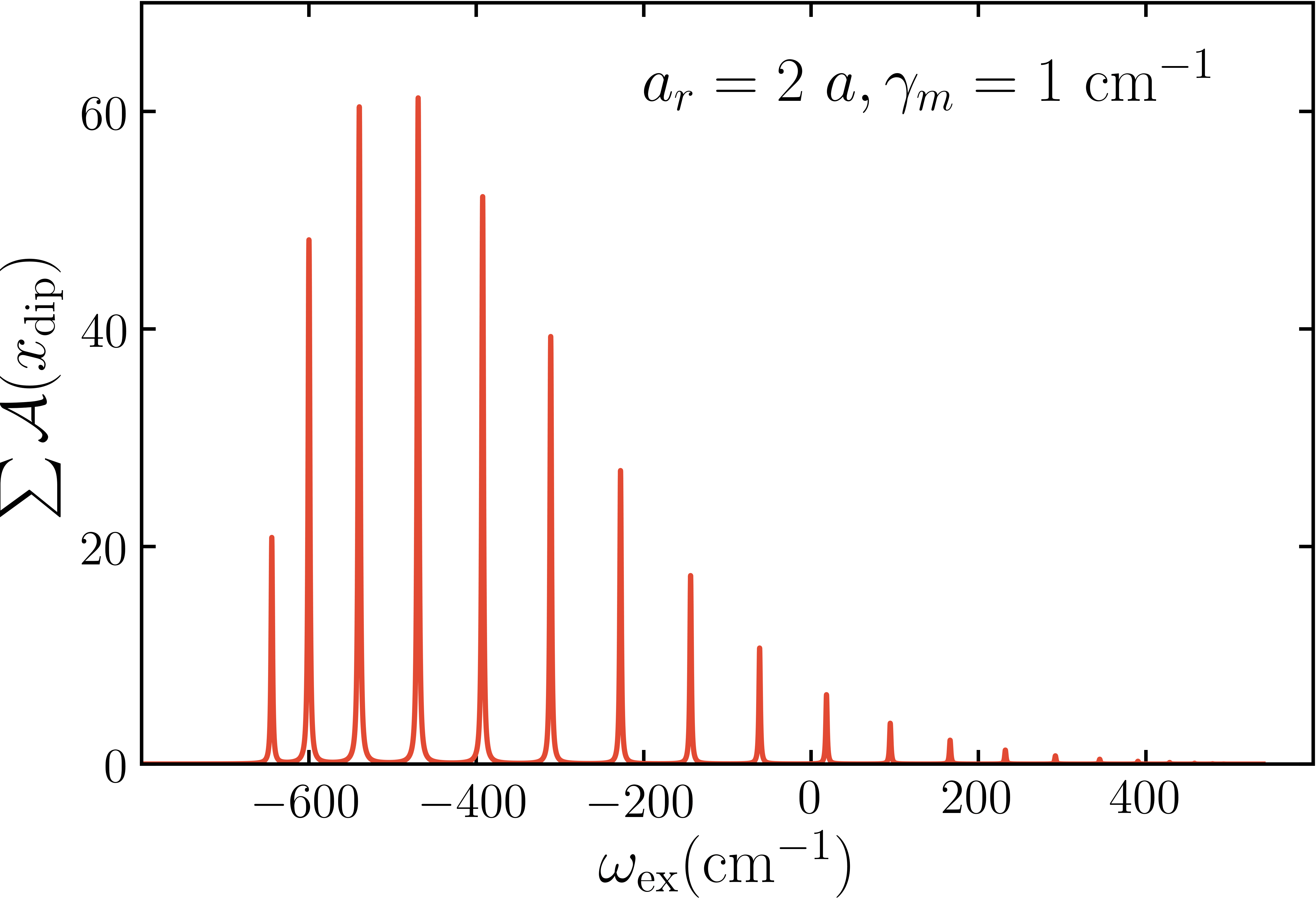}
   \includegraphics[width=14.5cm]{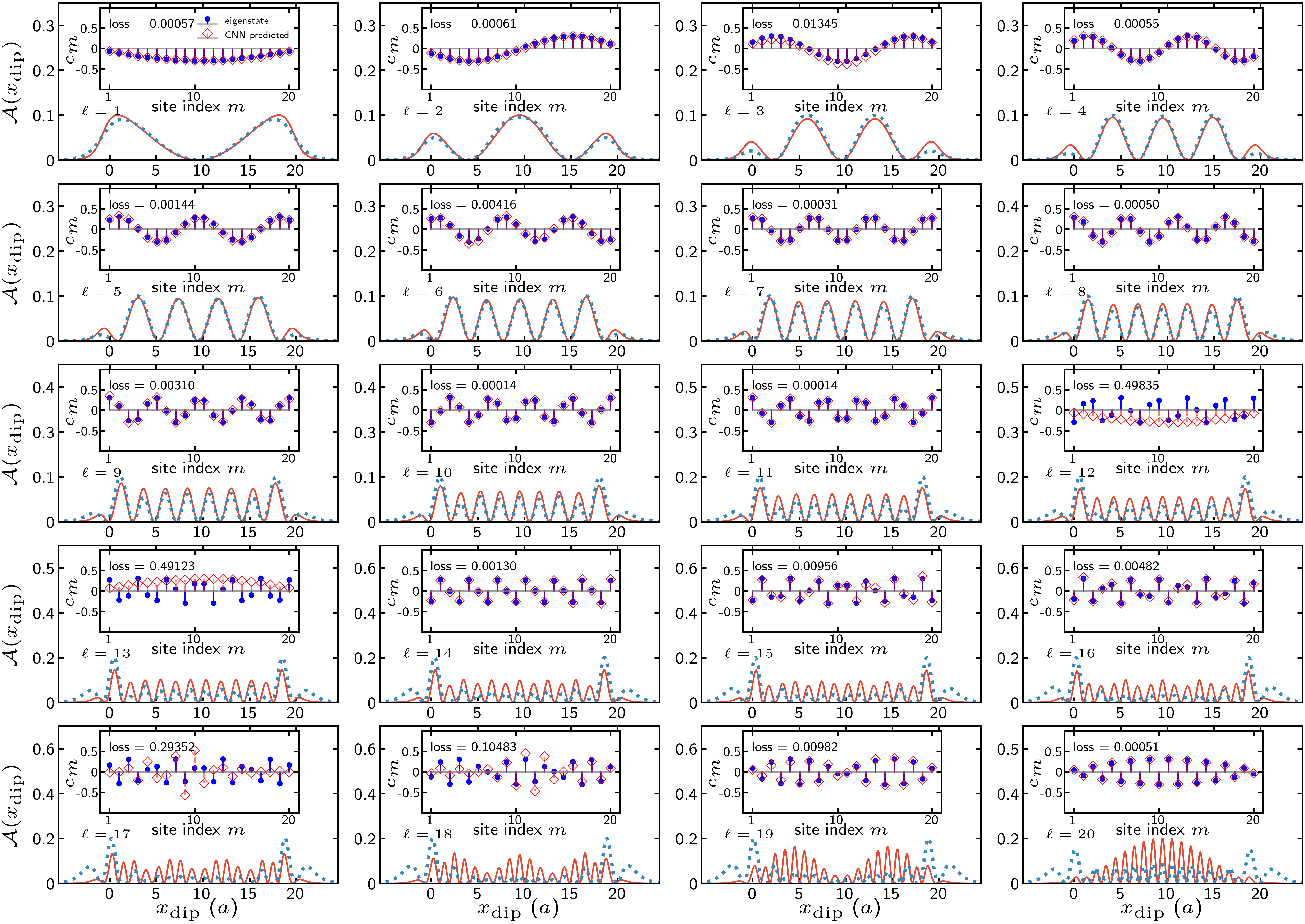}
   \caption{Top left: Frequency and spatial resolved spectrum.  Top right: Summed spectra over all tip-positions for each frequency. $a_{r} = 2a $, $\gamma_{m} = 1~\text{cm}^{-1}$, and distance between tip-edge and sample being $a$. To obtain explicit values we have taken $a=1.25\,\mathrm{nm}$. Bottom: Spatially resolved spectra at frequencies determined as described in the text. As comparison we show the result using Eqs.~(1)-(3) of the Letter. In the inserts we show the predicted wave function coefficients (red squares). Again, as comparison we show the eigenfunction (blue dots) from Eqs.~(1) and (2) of the Letter.}
  \label{fig:Polarizability_gamma1}
\end{figure}

\begin{figure}
  \centering 
  \includegraphics[width=6cm]{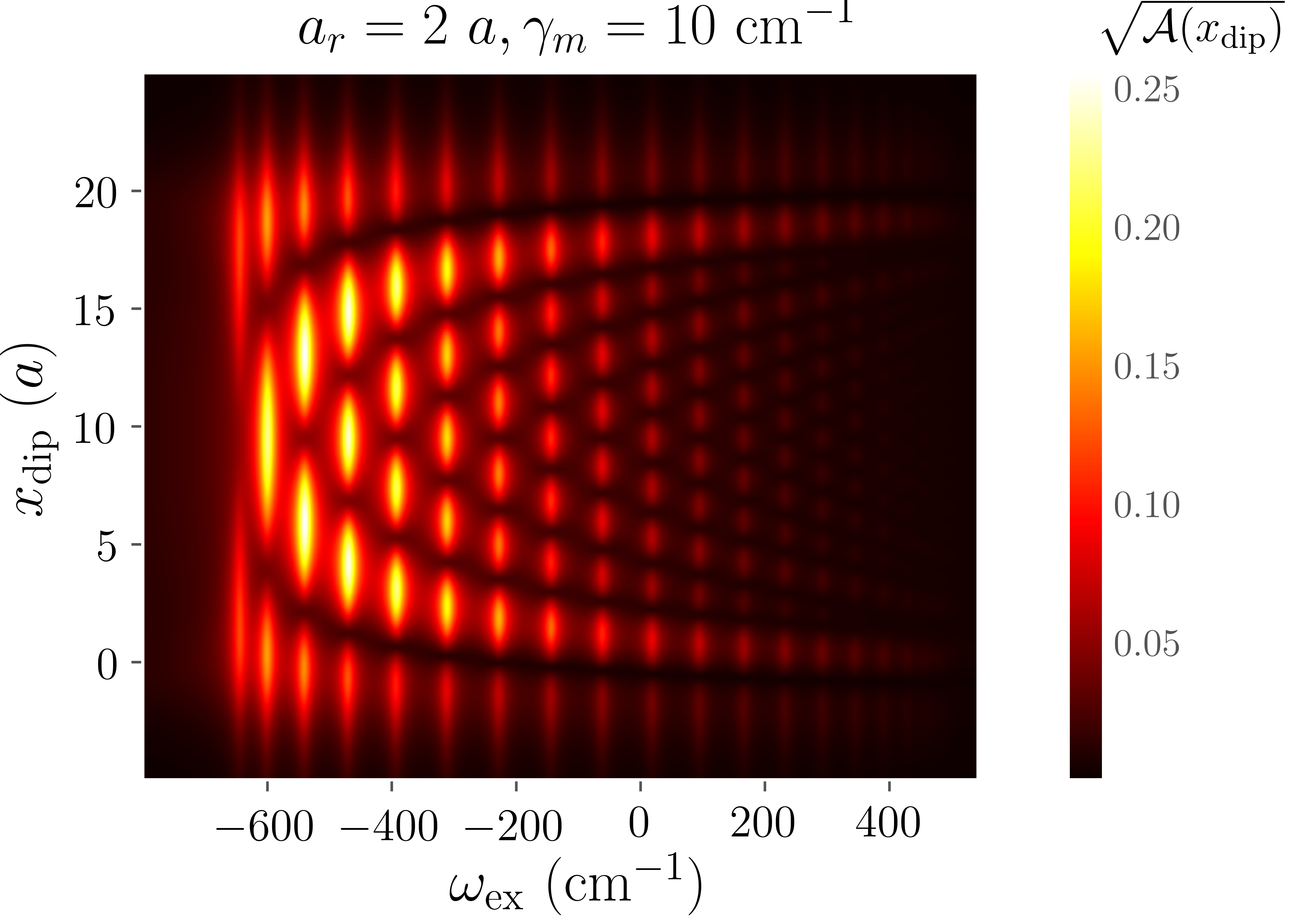}   
  \hspace{1cm}       
  \includegraphics[width=6cm]{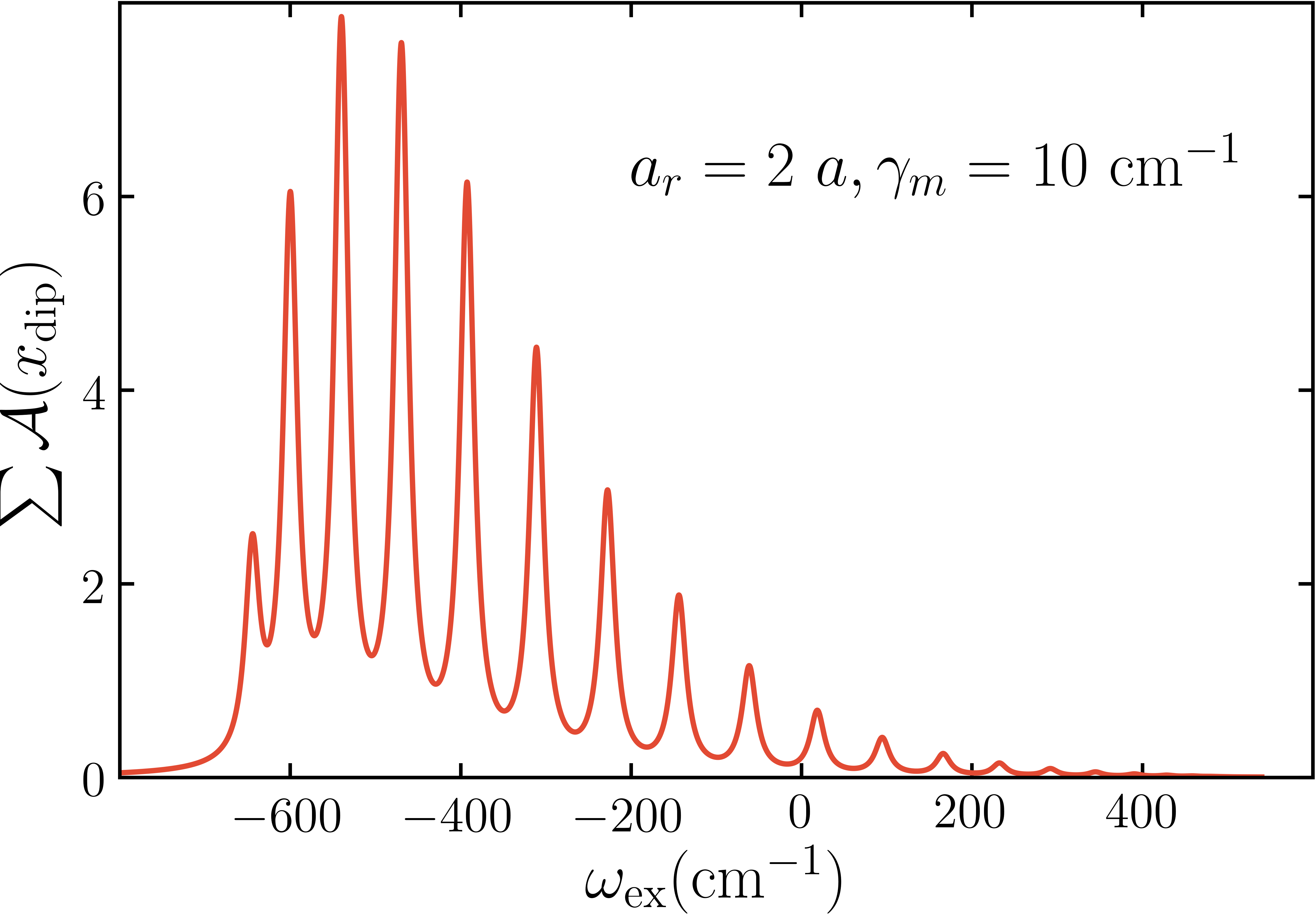}
  \includegraphics[width=14.5cm]{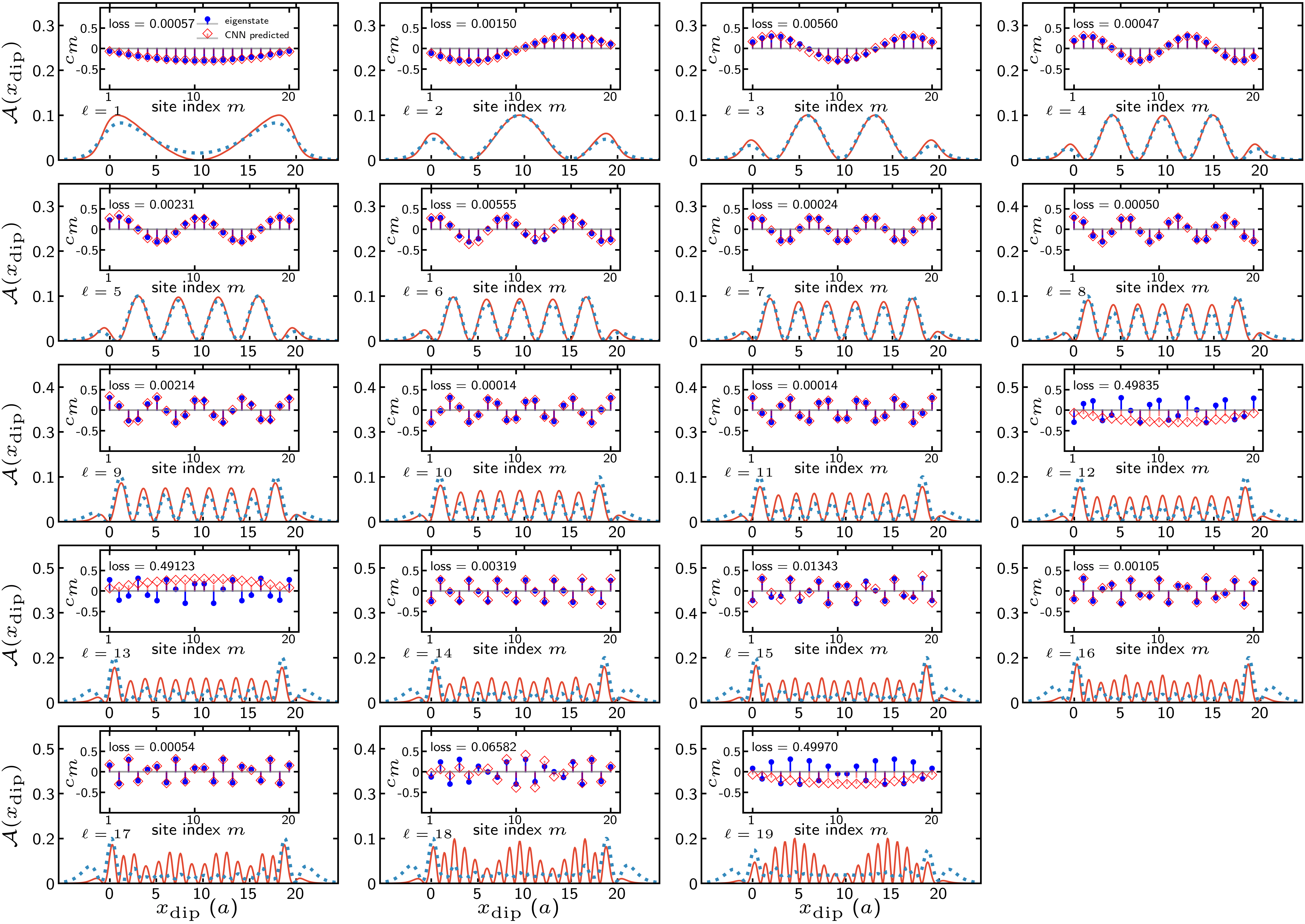}
  \caption{As Fig.~\ref{fig:Polarizability_gamma1} but for $\gamma_m=10\,\mathrm{cm}^{-1}$.}    
  \label{fig:Polarizability_gamma10}
\end{figure}

In Figs.~\ref{fig:Polarizability_gamma1} and \ref{fig:Polarizability_gamma10} we show in the top left frequency and spatially resolved spectra calculated using  the polarizability formalism for two different values of $\gamma_m$. In Fig.~\ref{fig:Polarizability_gamma1} we have a small value $\gamma_m=1\,\mathrm{cm}^{-1}$ and in Fig.~\ref{fig:Polarizability_gamma10} we have  $\gamma_m=10\,\mathrm{cm}^{-1}$. One clearly sees that in the case $\gamma_m=10\,\mathrm{cm}^{-1}$ there is considerable broadening, while the for the case $\gamma_m=1\,\mathrm{cm}^{-1}$ the peaks essentially do not overlap.

In principle we could investigate each frequency slice with the CNN. However, since we are interested in the eigenenergies, we adopt the following procedure: First we integrate over the spatial degree of freedom to obtain a spectrum that is now only a function of frequency (see top right plot). From this spectrum we determine the maxima of the peaks. For these frequencies we then consider the spatial dependence and use the corresponding spatial spectra as input to the CNN. If we cannot clearly identify a peak, the respective contribution is not taken into account. This happens in particular for high energies which correspond to eigenstates with many nodes of the wave function.
 
We see that even for quite large line-broadening $\gamma=10\,\mathrm{cm}^{-1}$ one can still quite well reconstruct the wave function coefficients corresponding to the Hamiltonian (1).

Note that in our calculations we have used a tip-aggregate distance that is equal to the distance $a$ between the molecules. For the results presented $a=1.25\,\mathrm{nm}$. This also implies that the Hertz dipole is located at $z_{\rm dip}=a+a_r=1.25\,\mathrm{nm}+2.5\,\mathrm{nm}=3.75\,\mathrm{nm}$. This distance is larger than the distance $z_{\rm dip}=2\,a$ used in the Letter. This implies that the rapid spatial variations are reduced in amplitude as can be seen in the ``cuts'' in Figs.~\ref{fig:Polarizability_gamma1} and Figs.~\ref{fig:Polarizability_gamma10}, in particular for high lying states which have many nodes of the wave function. In these plots we compare the ``cuts'' stemming from polarizability with the ideal results of the Letter (which are for a smaller distance $z_{\rm dip}$). Despite the considerable difference between these spectra (in particular for the high lying states) the reconstruction of the wave functions is remarkably accurate for most of the states. This is in particular surprising, since our CNN was trained with the ideal Hamiltonian (1) and with a smaller distance  $z_{\rm dip}$. These results indicate, that our adopted method is also robust against small variations in the tip-sample distance.

\clearpage
\newpage

\section{\label{sec:GaussianProcess} Gaussian process regression}
In addition to the neural network we have also tried to use Gaussian process regression \cite{RasmussenBook2006_SI} to reconstruct the eigenstate coefficients from near-field spectra. As in our previous work \cite{Bentley2018_SI} we used the optimization package MLOOP \cite{Wigley2016_SI}. Firstly, we obtain the eigenstate coefficients by diagonalizing the Hamiltonian and calculate the near-field spectra in turn. We set the maximum and minimum boundaries for the predicted coefficients to $1$ and $-1$, respectively. The predicted coefficients are normalized before they are used to calculate the near-field spectra. Then the optimization proceeds by minimizing the cost function, which is defined as the mean-squared-error between the near-field spectra calculated from the exact coefficients ($\mathcal{A}$) and that obtained from the predicted coefficients ($\mathcal{A}^{\rm pre}$), and is written as
\begin{equation}
{\rm Cost} = \frac{\sum_{i=1}^{N_{\rm tip}} \Big[\mathcal{A}(x_i)-\mathcal{A}^{\rm pre}(x_i)\Big]^2 }{N_{\rm tip}}
\end{equation}
with $N_{\rm tip} = 400$ referring to the number of positions at which the spectra is evaluated. We tried several optimization methods, including the Gaussian process method, the Nelder-Mead method, and the differential evolution method, which are all implemented in the MLOOP package. We set the maximum iteration number to 1000 and the target cost to $10^{-10}$. Finally the optimization will be terminated either the maximum iteration number is reached or the target cost is achieved. For a small system, e.g., $N=5$, all the wave function coefficients can be effectively predicted from the corresponding spectra. However, for a little larger system with $N=8$ molecules, wave functions of high energy eigenstates cannot be reproduced by any optimization approaches mentioned above. 



%

\end{document}